\documentclass[10pt,aps,pra,twocolumn,amsmath,amssymb,superscriptaddress]{revtex4-1}
\usepackage[english]{babel}
\usepackage[TS1,T1]{fontenc}
\usepackage{latexsym}
\usepackage{graphics}
\usepackage{epsfig}

\usepackage{color}
\usepackage{bm}
\usepackage{amsmath}
\usepackage{amssymb}
\usepackage{hyperref}
\graphicspath{{img/}} 
\usepackage{cancel}

\usepackage{textcomp}

\usepackage[normalem]{ulem}
\usepackage{color}
\usepackage{soul}

\newcommand{\bra}[1]{\langle #1|}
\newcommand{\ket}[1]{|#1\rangle}
\newcommand{\meanv}[1]{\left\langle #1 \right\rangle}
\newcommand{\bb}[1]{\left( #1 \right)}

\newcommand{\be}{\begin{equation}}
\newcommand{\ee}{\end{equation}}
\newcommand{\bea}{\begin{eqnarray}}
\newcommand{\eea}{\end{eqnarray}}

\newcommand{\oppsi}{\hat{\Psi}}

\definecolor{brown}{rgb}{0.59, 0.29, 0.0}

\newcommand{\1}{\uparrow}

\graphicspath{{img/}} 
\newcommand{\An}{\psi_{\rm Ansatz}}

\newcommand{\braket}[2]{\langle #1|#2\rangle}

\definecolor{green}{rgb}{0,0.75,0}

\usepackage[T1]{fontenc} 

\begin{document}
	
	\title{Dark solitons revealed in Lieb-Liniger eigenstates}
	
	\author{Weronika Golletz}
	\affiliation{Center for Theoretical Physics, Polish Academy of Sciences, Al. Lotnik\'{o}w 32/46, 02-668 Warsaw, Poland}
 	\affiliation{Instytut Fizyki Teoretycznej, Uniwersytet Jagiello\'nski, ulica Profesora Stanis\l{}awa \L{}ojasiewicza 11, PL-30-348 Krak\'ow, Poland}
	\affiliation{Department of Theoretical Physics, Faculty of Fundamental Problems of Technology, Wroc{\l}aw University of Science and Technology, 50-370 Wroc{\l}aw, Poland}
	
	\author{Wojciech~G{\'{o}}recki}
	\affiliation{Center for Theoretical Physics, Polish Academy of Sciences, Al. Lotnik\'{o}w 32/46, 02-668 Warsaw, Poland}
	
	\author{Rafa\l~O{\l}dziejewski}
	\affiliation{Center for Theoretical Physics, Polish Academy of Sciences, Al. Lotnik\'{o}w 32/46, 02-668 Warsaw, Poland}
	
	\author{Krzysztof~Paw{\l}owski}
	\affiliation{Center for Theoretical Physics, Polish Academy of Sciences, Al. Lotnik\'{o}w 32/46, 02-668 Warsaw, Poland}

	
	\begin{abstract}
We study how dark solitons, i.e. solutions of one-dimensional single-particle nonlinear time-dependent Schr\"odinger equation,
emerge from eigenstates of a linear many-body model of  contact interacting bosons  moving on a ring, the Lieb-Liniger model. This long-standing problem was addressed by various groups, which presented different, seemingly unrelated, procedures to reveal the solitonic waves directly from the many-body model. 
Here, we propose a unification of these results using a simple Ansatz for the many-body eigenstate of the Lieb-Liniger model, which gives us access to systems of hundreds of atoms. In this approach, mean-field solitons emerge in a single-particle density through repeated measurements of particle positions in the Ansatz state. The post-measurement state turns out to be a wave packet of yrast states of the reduced system.


\end{abstract}

	
	
	\maketitle

\section{Introduction}

The famous Lieb-Liniger (LL) model~\cite{Lieb1963, LiebLiniger1963} describes particles moving along a circle and interacting via delta inter-atomic potential. Such a simple interaction turns out to be a well-suited approximation for realistic interactions between neutral slow atoms. Thus, the LL model  and its extensions, remain an active research topic in theoretical and experimental physics and mathematics~\cite{Gogolin2004Dec, Jiang_2015,Lang2017,Cazalilla2011, Sowinski2019Sep}.
	
	The same system, of $N$ atoms with contact interaction, is often treated within a simple mean-field (MF) approximation, based on the non-linear Schr\"odinger  equation (NLSE):
	\begin{equation}
	i\hbar\,\partial_t \phi_{\rm MF}(x, t) = \left(-\frac{\hbar^2 \partial_x^2 }{2m}+ g (N\!-\!1) |\phi_{\rm MF} (x,t)|^2 \right) \phi_{\rm MF} (x,t), 
	\label{eq:NLSE}
	\end{equation} 
	where the wave function $\phi_{\rm MF}(x, t)$ is interpreted as an orbital occupied by a macroscopic number of atoms, $m$ is the particle mass, and $g$ is the  interaction strength. 
	The latter equation~\eqref{eq:NLSE},  is useful in many areas of physics ranging from quantum optics \cite{Kivshar1998May} to hydrodynamics~\cite{Gross1961, Johnson1976}.
	It is also a rare example of a model with physical applications supporting solitonic solutions \cite{zakharov73}, observed
	in atomic gases \cite{Burger1999}, plasma \cite{heidman2009}, water waves \cite{Chabchoub2013}, ferromagnetic materials \cite{Tong2010}.
	In the case of gases, when their atoms repel each other, i.e. $g>0$, a soliton is a rarefaction in the atomic density, which moves with a constant speed, preserves its shape and is unusually robust thanks to the balance between dispersion and nonlinearity~\cite{Frantzeskakis2010}. 
	In this case ($g>0$) the soliton is called a dark soliton, which can be either black or gray. Black solitons are characterized by a zero-density dip i.e. a point where the atomic density is exactly zero and the phase of $\phi_{\rm MF}(x)$ undergoes a sharp $\pi$-jump. While gray solitons have a non-zero density dip with the phase jump strictly smaller than $\pi$.
	
\begin{center}
		 \begin{figure}[h!]
			\includegraphics[]{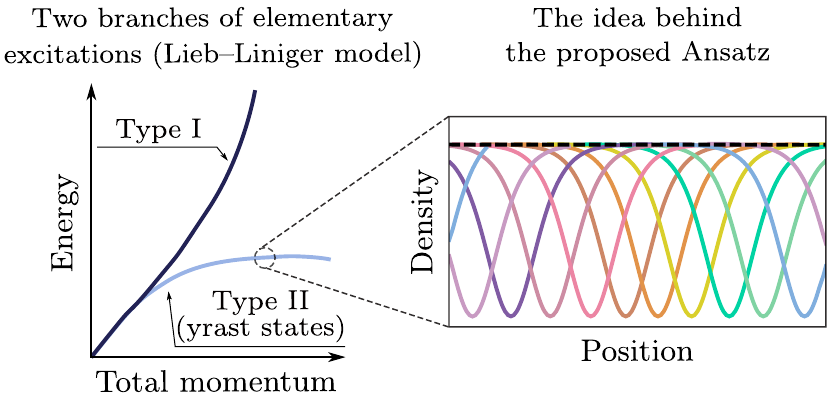}
			\caption{(color online)
			As shown in Ref.~\cite{Lieb1963}, among eigenstates of the Lieb-Liniger model there are special states, forming two branches of the many-body elementary excitations (left panel). 
			In this paper we show that states forming the type-II spectrum, 
			i.e. yrast states, can be approximated by a superposition of the mean-field dark solitons with relative phases (marked with color) depending on the solitons' positions (right panel). We discuss the validity of a suitable Ansatz and use it to unify different views on  correspondence between the yrast states of the Lieb-Liniger model and the mean-field solitons from the non-linear Schr\"{o}dinger equation. 
			\label{fig:sol-sketch}}
		\end{figure}
	\end{center}

	There is a puzzling link between the MF dark solitons and the solutions of the underlying many-body LL model. More than a decade after the seminal paper by E. Lieb~\cite{Lieb1963}, a coincidence between the dispersion relations of dark solitons and certain many-body eigenstates, the so-called type-II elementary excitations,  was observed in the weak interaction limit~\cite{Kulish1976,ishikawa1980}. The type-II excitations are simply the many-body eigenstates that minimize the energy for a fixed total momentum, sometimes called yrast states~\cite{Mottelson1999}. Together with the
	 type-I excitations (corresponding to  Bogoliubov quasiparticles~\cite{Lieb1963})  they constitute two branches of elementary excitations, with dispersion relations sketched in the left panel of Fig.~\ref{fig:sol-sketch}.
	

Why is the correspondence between the yrast states and dark solitons bizarre? 
Firstly, yrast states, as eigenstates of the system, are stationary solutions, whereas the MF dark solitons are solutions of time-dependent Eq. \eqref{eq:NLSE}.
Moreover, the LL model includes all correlations between particles in a linear Hamiltonian, while the MF approach gets rid of mutual correlations but introduces the non-linearity in the description. Finally, the eigenstates of the many-body model have to be translationally invariant. On the other hand, the dark solitons do not obey this symmetry.
As the nature of the MF solitons and the yrast states is so different, the question arises whether they have something in common except the same dispersion relation.

There have been efforts to show, that the relation between these objects is deeper, and in particular that the MF soliton can be extracted directly from the yrast states.
In Refs.~\cite{syrwid2015, syrwid2016} it was indicated that MF solitons are already hidden in a single yrast state, and they will emerge in  sufficiently high order correlation functions.
In turn, in Refs.~\cite{Sato2012, Sato2016, Kaminishi2019, Brand2019} it was shown 
that MF soliton appears in a single-particle density, when calculated for an appropriate superposition of many yrast states.
The other relations between the many-body states and solitons were also presented in~\cite{martin2010prl, Katsimiga2017bent, Katsimiga2018, Kaminishi2019, Delande2014}. 
Still, these interesting results leave the field in an unpleasant situation of many seemingly unrelated views on the connection between MF solitons and many-body yrast states.

Here, we unify different approaches by employing a simple but powerful Ansatz for the yrast state of the LL model in the MF regime of parameters.
We use this Ansatz, to show that the state which appears after "measuring" many particles drawn from a high order correlation function is a random superposition of yrast states. The mutual unification between two approaches becomes  apparent
as we consider systems consisting of hundreds of atoms close to the MF regime.

The paper is organized as follows. 
In Section~\ref{sec:ll} we remind LL model, yrast state and define the parameter regime we are interested in.
Our Ansatz for yrast states is introduced in Section~\ref{sec:ansatz}. Validity of this Ansatz is discussed in Section~\ref{sec:validity}. In Sections~\ref{sec:highorder} and~\ref{sec:superposition} we show how the different constructions of the MF solitons out of the yrast states presented in Ref.~\cite{syrwid2015} and Refs.~\cite{Sato2012, Sato2016, Kaminishi2019} prove to be different views of the same object.
To make this paper self-complete we describe in details all relevant analytical previous results and our numerical approaches in Appendices.

\section{The Lieb-Liniger model and yrast states\label{sec:ll}}
We will investigate eigenstates of the many-body system of $N$ bosons moving along a circle of length $L$ governed by the LL Hamiltonian \footnote{We rescaled the original Lieb-Liniger model by a factor $2$, to have the form which is used more frequently now.}:
	\begin{equation}
	\hat{H} = -\frac{\hbar^2}{2m} \sum_{j=1}^{N} \partial_{x_j}^2 + g \sum_{\substack{j,\, l \\ j<\,l}}^N \delta(x_j - x_l)\,,
	\label{eq:LLham}
	\end{equation}
where $x_j$ denotes position of $j$-th particle. 
As this system is translationally invariant, the values of total momentum $\hat{P} = -i\hbar\,\sum_{j=1}^N\partial_{x_j}$ can be used to label the energy eigenstates, even in the case with interaction.
The exact solutions for the eigenstates of the LL Hamiltonian \eqref{eq:LLham} for repelling ($g>0$)  particles are known since  1963~\cite{Lieb1963}. Among the eigenstates, there are special ones that are called elementary excitations, which can be divided into two families. Before Ref. \cite{Lieb1963}, 
the approximated theories was applied to find energies of weak perturbations of an atomic gas. This originated in a single family of the Bogoliubov elementary excitations (identified with the type-I excitations). Its dispersion relation given by $E_{\rm B } (P) = \sqrt{\frac{P^2}{2 m} \bb{\frac{P^2}{2 m} + \frac{2 g N}{L} } }$ \cite{deWitt1958}.

The unexpected second family, revelead by the exact solution presented in \cite{Lieb1963}, consists of the aforementioned yrast states. These are also called "one hole excitations" \cite{Lieb1963},  or the "lowest energy solutions for fixed total momentum" \cite{Mottelson1999}. The yrast state with the total momentum of $P=\frac{2\pi\hbar}{L} K$ is represented by  $\ket{K}$, where $K$ is an integer.	

As the subject of this paper is the relationship between the yrast states and solitons, we 
will restrict our considerations to MF regime  as the NLSE should work, in principle, only there. 
That is, within the regime of weak interactions with only slightly correlated atoms, in which quantum phenomena, like the quantum depletion of the ground state, are small. On the other hand, it is desired to see the effects that are substantially different from the ideal gas case. Therefore, we require the healing length $\xi := \sqrt{\hbar^2L/ g m N}$,
which is close to the soliton width, be much shorter than the system size $L$ \cite{rafal2018idealSol}. If we were to consider a small number of atoms, the latter condition would lead to a large $g$, resulting in strong interactions. Therefore, we are interested in the limit in which number of atoms $N$ converges to infinity, an interaction strength $g$ goes to $0$, but the MF parameter $ng := \frac{\hbar^2}{m \xi^2}$, with $n:=N/L$ denoting a gas density, is fixed.
In this limit the LL coupling constant $\gamma\propto g/N $  \cite{Lieb1963} decreases with the number of atoms as $(ng)/N^2$, indicating that indeed the system enters quickly a weak interaction regime.

The MF regime defined in such a way is very difficult to handle in the frame of many-body analysis, which is usually limited to systems with small number of atoms $N$. 
Apart from the few existing semi-analytical results~\cite{Sato2012, Brand2019, Kaminishi2019}, the majority of approaches are devoted to small systems of $\approx 10-20$ atoms~\cite{KanamotoCarr2008, Kaminishi2011, Fialko2012, syrwid2015, syrwid2016, rafal2018roton} solved with brute force methods or around $\approx 100$ atoms solved with sophisticated and time consuming numerics~\cite{Katsimiga2017db, Delande2014, Mistakidis2018, Katsimiga2018}. 
Our way around these numerical difficulties is to use a natural and simple Ansatz for the yrast states in the MF regime. 

\section{The Ansatz for yrast states}\label{sec:ansatz}
Here we shall discuss, step-by-step, our construction of the Ansatz for the yrast states in the MF regime.
The main building block of the Ansatz consists of a product state of $N$ particles occupying a single orbital $\phi(x)$ represented by $ \prod_{j=1}^N\phi (x_j)$. Next, the Ansatz has to belong to the same momentum space as the yrast state $\ket{K}$, i.e. the translation of the $N$-particle wave function by $\Delta x$ needs to be equivalent to multiplying it by a factor $e^{i\frac{2\pi}{L}K\Delta x}$:
\begin{equation}
    \psi(x_1+\Delta x,...,x_N+\Delta x)=e^{i\frac{2\pi}{L}K\Delta x}\psi(x_1,...,x_N).
\end{equation}
For any orbital $\phi$ one may construct the states satisfying above condition by taking a continuous superposition of product states shifted by the translation operator $e^{-i \hat{P} y/\hbar}$ and multiplied by the phase factor $e^{i\frac{2\pi}{L}Ky}$ over all possible shifts $y$ (see Appendix \ref{appendix:ansatz-momentum} for formal justifications):
\begin{equation}
	\psi (x_1,\ldots, x_N) = \mathcal{N} \int_0^L\,dy\,e^{i \frac{2\pi}{L}K y} \prod_{j=1}^N \phi(x_{j}-y),
	\label{eq:smeared-sol-cos}
\end{equation}
where $\mathcal{N}$ is a (real) normalization factor.

As a yrast state  $\ket{K}$ is the lowest energy state for fixed total momentum equal exactly to $\frac{2\pi\hbar K}{L}$, therefore the task would be to find  an orbital ${\phi}$, that minimizes the average energy of the state \eqref{eq:smeared-sol-cos}.
Finding such orbital would be a  difficult task, as the average energy of the state \eqref{eq:smeared-sol-cos} is given by a complicated formula (see Appendix \ref{appendix:energy}).
On the other hand it is known that energies of the yrast states $\ket{K}$ and  MF solitons agree \cite{Kulish1976}. Therefore, as the Ansatz for yrast state we choose the state \eqref{eq:smeared-sol-cos} with 
${\phi}(x) = {\phi}_{\rm MF}(x)$, where ${\phi}_{\rm MF}(x-v t)$ is the solitonic solution of the NLSE with the average single particle  momentum $\meanv{-i\hbar\,\partial_x}$ equal to $2\pi\hbar \,K/ (NL)$:
\begin{equation}
	\psi_{\rm Ansatz}(x_1,\ldots, x_N) = \mathcal{N} \int_0^L\,dy\,e^{i \frac{2\pi}{L}K y} \prod_{j=1}^N \phi_{\rm MF}(x_{j}-y)
	\label{eq:smeared-solv2}
\end{equation}
We also use state defined in Eq.  \eqref{eq:smeared-solv2} in the Dirac notation:
	\begin{equation}
	\ket{\psi_{\rm Ansatz}} =\mathcal{N} \int_0^L\,dy\,e^{i \frac{2\pi}{L}K y} e^{-i \hat{P} y /\hbar} \ket{\phi_{\rm MF}}^{\otimes N}.
	\label{eq:smeared-sol}
\end{equation}

The exact form of solitonic solution on the circle ${\phi}_{\rm MF}$ is quite complicated -- we give the appropriate formulas and our numerical methods for handling them, in Appendix \ref{appendix:mean-field}.


The construction of the Ansatz is sketched in Fig. \ref{fig:sol-sketch} -- the many-body Ansatz is understood as a continuous superposition of macroscopically occupied MF solitons. Each soliton in the superposition \eqref{eq:smeared-solv2} appears with a phase prefactor $e^{i 2 \pi K y / L}$ (distinguished in  Fig. \ref{fig:sol-sketch} with a color) depending on the position shift $y$.

We remark that the Ansatz follows the ideas partially spread in the  community, that the yrast states 
should be somehow related to the product states of MF solitons but with unknown position of the density dips, i.e. smeared over the whole circle as shown in Fig.~\ref{fig:sol-sketch}.
Such Ansatz was presented in the context of Bose-Einstein condensation \cite{Castin2001}. Condensate, as defined via Penrose-Onsager criterion \cite{Penrose1956}, is supposed to appear in the system of bosons at very low temperature. As show in Ref. \cite{Castin2001}, surprisingly, even at $T=0$, when the system is in a ground state, there may be no condensation at all. This happens when the system has some continuous symmetry, like the translational invariance in our case. 
Once the symmetry is broken, for instance by measuring positions of few bosons, condensation may emerge immediately. The situation presented in this paper is similar but the resulting condensate is (i) not a ground state of the system and (ii) is {\it temporary} as it may disappear in time \cite{syrwid2016, Sato2016}.

\begin{center}
	\begin{figure}[h!]
		\includegraphics[]{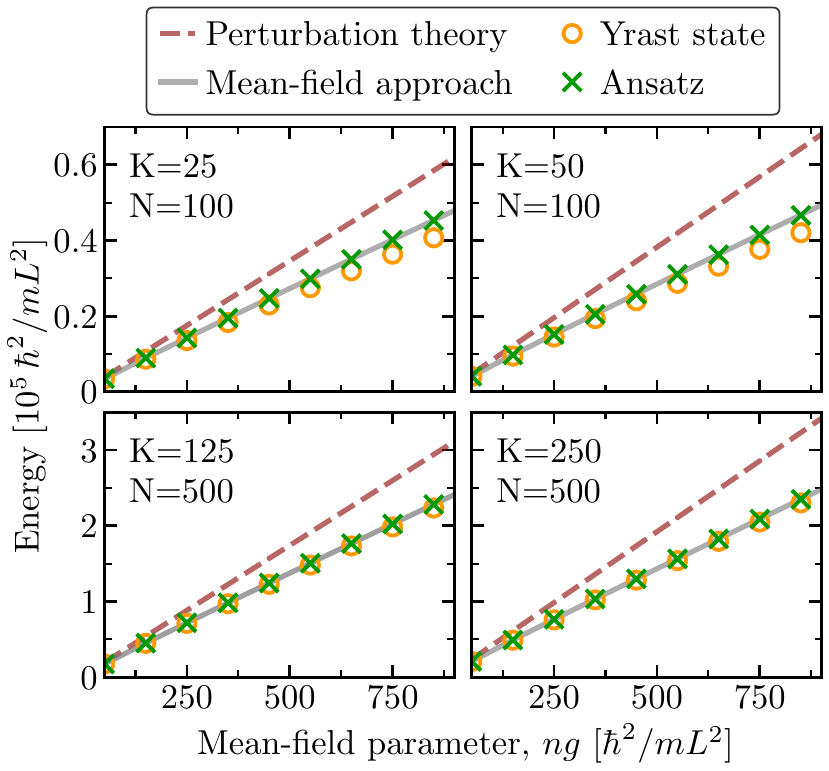}
		
		\caption{(color online) Expectation value of Lieb-Liniger Hamiltonian~\eqref{eq:hamLL-2-quant} evaluated as a function of the mean-field parameter $ng$ in the mean-field approximation, solution of Eq. ~\eqref{eq:NLSE} (solid gray line), for the yrast state~(Appendix \ref{appendix:LLeqs}) (empty yellow circles), for the Ansatz~\eqref{eq:smeared-sol} (green crosses), and 	using perturbation theory (dashed burgundy line). In the top panels $N=100$, while in the bottom ones $N=500$. The parameter $K$ is set to $N/4$ in the panels to the left and to $N/2$ in the panels to the right, as indicated in the upper left corner of each graph. The top panels share a common energy scale and so do the bottom ones.}
        \label{fig:energies}

	\end{figure}
\end{center}

How accurately does the energy of the Ansatz agree with the energy of the exact LL solution, the yrast state, as the interaction strength increasing? 
In Fig. \ref{fig:energies}, we present the energies as functions of the MF parameter $ng$, where $n=N/L$ is the gas density, for the yrast state with $K=N/2$ (black soliton) and $K=N/4$ (gray soliton) and the corresponding Ansatz \eqref{eq:smeared-sol} (see Appendices \ref{appendix:energy}, \ref{appendix:numerical-methods} and \ref{appendix:LLeqs} for the details of computations). 
For the reference, we plot results obtained via the first order perturbation theory with interaction strength $g$ being a small parameter (dashed burgundy line). In this case the average energy is evaluated in the yrast state corresponding to $g=0$, which is a state
with $N-K$ motionless particles and the remaining $K$ of them with momentum $2\pi \hbar/L$.
Its average energy is a linear function of interaction strength $g$ equal to $K\frac{4\pi^2\hbar^2}{2 m L^2} +\frac{g}{2L} \bb{N(N-1)+2NK-2K^2}$. 
The second reference curve, is the average energy of the MF state, with $N$ atoms occupying a single orbital $\phi_{\rm MF}$ (solid gray line). 

As expected the energy of yrast state, MF state, and Ansatz are close to each other, even for such strong interactions that first order perturbation theory fails. 
Moreover, in the ideal gas limit, our Ansatz \eqref{eq:smeared-sol} is exactly equal to the yrast state	$\lim_{g\to0}\ket{\psi_{\rm Ansatz}} \equiv \lim_{g\to0}\ket{K}$ for any $K\neq 0$ \cite{rafal2018idealSol}. In the same limit, the energy of the dark MF soliton is actually slightly smaller than the energy of the yrast state  (see Appendix \ref{appendix:ideal-gas}).

The intuitive definition of an Ansatz, together with the apparent agreement between its energy and the energy of yrast state, motivate us to use the Ansatz \eqref{eq:smeared-sol} instead of yrast state in our study on the correspondence between LL model and NLSE. 
Before we shall do it, we discuss in more detail the validity range of  the Ansatz.


\section{Validity range}\label{sec:validity}

We analyse the validity range by determination how well our Ansatz approximates the true yrast state for a given strength of interaction. The most objective and unambiguous measure of the similarity between two states is fidelity $|\braket{\psi_{\rm Ansatz}}{K}|^2$. However, to calculate it directly one would need to express the yrast state in the position representation (or to express the Ansatz in terms of quasi momenta from solutions of the LL model), which would be an extremely demanding task. Therefore we propose a simple (but very rough) lower bound for the value of fidelity, based on the values of energies calculated independently in each of the formalisms mentioned.

\begin{center}
	\begin{figure}[h!]
		\includegraphics[]{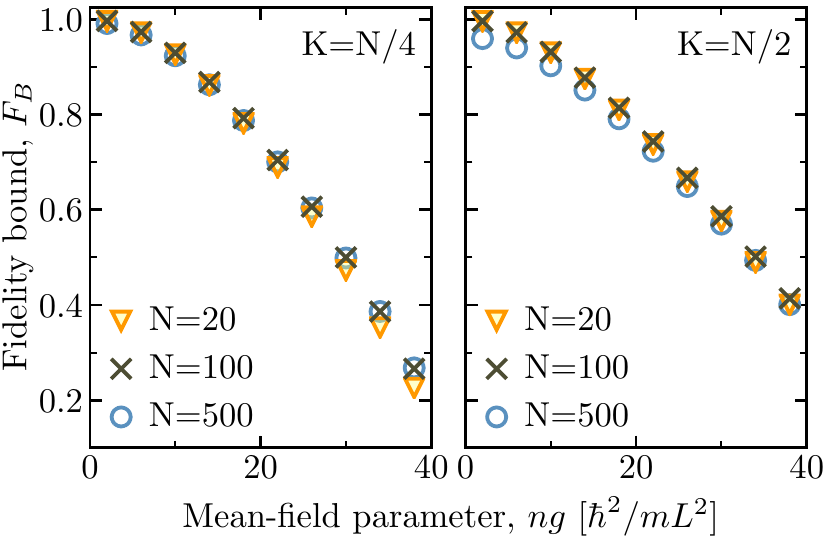}
		\caption{(color online) A rough lower bound for the fidelity between the Ansatz~\eqref{eq:smeared-sol} and yrast state, $F_{B}$~\eqref{FB}, as a function of the mean-field parameter $ng$ obtained for $N=20\,\text{(triangles)},\, 100\,\text{(crosses)},\, 500\,\text{(circles)}$. The parameter $K$ is set to $N/4$ in the panel to the left and to $N/2$ in the panel to the right, as indicated in the upper right corner of each graph. The panels share common vertical scale.
		\label{fig:validity}}
	\end{figure}
\end{center}

As the Ansatz has a well-defined total momentum, therefore it can be decomposed into the basis of many-body eigenstates of the same total momentum:
\begin{equation}
\ket{\An} = \alpha_0 \,  \ket{K} + \sum_{j=1}^{\infty} \alpha_j\,\ket{K^j},
\label{eq:An-deco}
\end{equation}
where $\ket{K^j}$ is the $j$-th excited eigenstate with total momentum $\frac{2\pi\hbar}{L}K$ and energy $E_{\rm exc}^{j}$, and $\sum_{j=0}^\infty |\alpha_j|^2=1$.
The excited eigenstates are listed in the ascending order with respect to their energy, i.e. $E_{\rm exc}^1 \leq E_{\rm exc}^2 \leq E_{\rm exc}^3 \leq \ldots $.

The average energy $\bra{\An} \hat{H} \ket{\An}$ can be expressed with the help of Eq. \eqref{eq:An-deco} as:
\begin{eqnarray}
E_{\rm Ansatz} &=& |\alpha_0|^2 E_{\rm yrast} + \sum_{i=1}^{\infty} |\alpha_i|^2\,E_{\rm exc}^i \\
  &\geq& E_{\rm yrast} + \bb{1-|\alpha_0|^2 }\,\bb{E_{\rm exc}^1 - E_{\rm yrast}} \nonumber,
\end{eqnarray}
where we use an inequality $E_{\rm exc}^j > E_{\rm exc}^1$ and $E_{\rm yrast}$ is the energy of the yrast state with the total momentum $\frac{2\pi\hbar}{L}K$.
Therefore, the fidelity between the Ansatz \eqref{eq:smeared-sol} and the corresponding yrast state, i.e. $|\alpha_0|^2$, obeys the inequality
\begin{equation}
|\braket{\An}{K}|^2\geq \frac{E_{\rm exc}^1 - E_{\rm Ansatz} }{E_{\rm exc}^1 - E_{\rm yrast}}=:F_{B},
\label{FB}
\end{equation}
where we introduced a fidelity bound $F_B$. 

In Fig. \ref{fig:validity} we show how $F_B$ decreases with increasing interaction strength. 
First of all, we observe, that data points calculated for a different number of atoms, but for the same value of mean-field parameter  $ n g$, are close to each other. 
Second, we see that for small values of $ng$ the fidelity between the Ansatz and the yrast state has to be very high (close to one).  
Here, it is worth stressing the fact, that the relatively small value of $F_B$ for much bigger $ng$ does not automatically implicate uselessness of the Ansatz for stronger interactions -- the fidelity may be still close to one, but calculating it directly would require much more advanced numerical methods and would be unfeasible for large number of atoms.


Given that for $ng \lesssim 25$ the Ansatz is a good approximation for the yrast state, we proceed to use it as a replacement for the yrast state to unify the results of other groups. We will benefit from the fact, that for the Ansatz many calculations may be done analytically  and the remaining necessary numerical analysis is feasible even for large number of atoms.

\section{Dark solitons revealed in high order correlation functions}\label{sec:highorder}
In this section, we study the emergence of dark solitons out of a single yrast state  as it was done in Ref. \cite{syrwid2015}. 
To make the comparisons with the literature results easier, we use  the second quantization formalism, in which the energy operator \eqref{eq:LLham} may be written as
\begin{equation}
\hat{H} = \int_0^L \text{d}x\,  \bigg[\frac{-\hbar^2}{2m}\oppsi^{\dagger} (x) \partial_x^2 \oppsi(x) + \frac{g}{2} \oppsi^{\dagger}(x)^2\oppsi(x)^2\bigg],
\label{eq:hamLL-2-quant}
\end{equation}
where $\oppsi(x)$ ($\oppsi^{\dagger}(x)$) is a annihilation (creation) field operator of a boson at position $x$  satisfying the commutation relations $[\oppsi(x), \oppsi^{\dagger}(x')] = \delta(x - x'),\, [\oppsi(x), \oppsi(x')] = [\oppsi^{\dagger}(x), \oppsi^{\dagger}(x')] = 0$. 
The second quantization formalism is also very handy in performing any computations within the Ansatz (see Appendix \ref{appendix:ansatz}).

The object of interest in Ref. \cite{syrwid2015} is a $m$-th order correlation function
\begin{align}
&\rho_m(x) \propto \label{eq:mordercor} \\ 
&\bra{K} \oppsi^{\dagger}(x_1)...\oppsi^{\dagger}(x_{m-1})\oppsi^{\dagger}(x)\oppsi(x)\oppsi(x_{m-1})...\oppsi(x_1) \ket{K}\nonumber,
\end{align}
normalized to $1$.
From its mathematical structure, Eq.~\eqref{eq:mordercor} is identified as a probability density function (PDF) from which one draws a random position $x_m$ of the $m$-th particle to be measured.
In Ref.~\cite{syrwid2015}, function~\eqref{eq:mordercor} is considered for increasing $m$ after subsequent "measurements" of particles. It was observed that $\rho_m$ resembles density $|\phi_{\rm MF}|^2$ of the MF soliton from the NLSE~\eqref{eq:NLSE}.
 However, in Ref.~\cite{syrwid2015} the agreement between MF solitons and the $m$-th order correlation function of a yrast state was demonstrated only for the healing length $\xi > L$ and for a few particles, namely for almost a non-interacting system. Contrarily, in the present paper, the Ansatz enables the study of healing lengths much smaller than the size of the system $L$ for a large number of particles. 
 To that end, we employ the following procedure using Eq.~\eqref{eq:mordercor} with the yrast state $\ket{K}$ replaced by our Ansatz $\ket{\psi_{\rm Ansatz}}$~\eqref{eq:smeared-sol}. We begin with a computation of the single-particle reduced density matrix $\rho_1$ which is used as a PDF to draw a random position $x_1$. Subsequently, we compute second-order correlation function $\rho_2$ which again serves as the PDF for the next random position $x_2$ draw. Repeating this process $m-1$ times outputs the $m-1$ positions,  parameters of the marginal distribution $\rho_m(x)$ \eqref{eq:mordercor}, evaluated for the Ansatz~\eqref{eq:smeared-sol} (for details of our calculations see Appendix \ref{appendix:computation-state}).

\begin{center}
		\begin{figure}[h!]
			\includegraphics[]{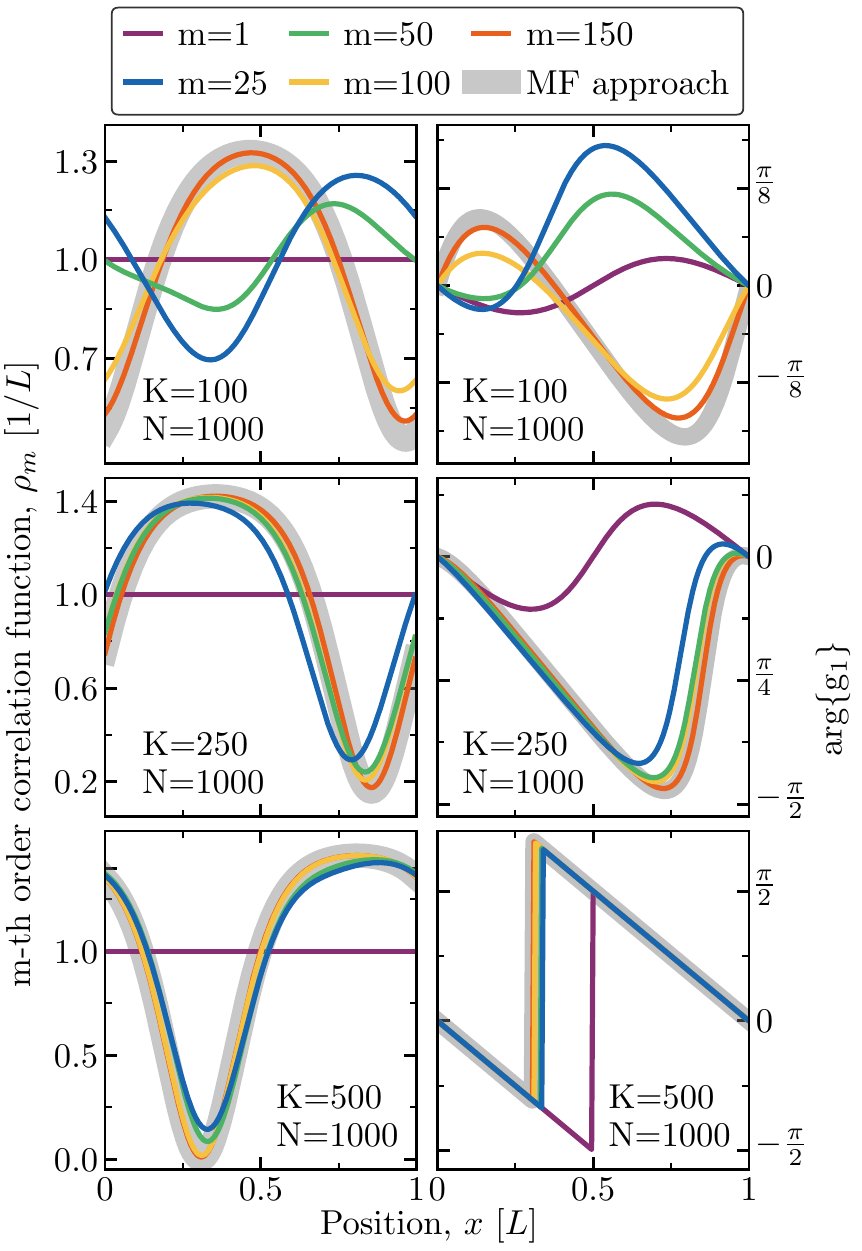}
 			\caption{(color online) Left: The $m$-th order correlation functions $\rho_m$~\eqref{eq:mordercor}. Right: Phase ${\rm arg}\left\{{\rm g}_1\,\right\}$ of the correlation function~\eqref{eg:g1}. Both plotted with respect to the position of $m$-th particle $x$. Different solid lines correspond to different orders $m$ from the shallowest to the deepest dip in the left panels $m = 1\,\text{(violet)},$ $25\,\text{(blue)},$ $50\,\text{(green)},$ $100\,\text{(yellow)},$ $150\,\text{(red)}$. The thick gray line corresponds to the mean-field (MF) solitonic solution of Eq.~\eqref{eq:NLSE}. Parameters: $ng=25,$ $N=500,$ and $K = 50,$ $125,$ $250$ for top, middle and bottom rows respectively, as indicated in the bottom-right corner of each graph.}

\label{fig:density}
					\end{figure}
	\end{center}

In the left panel of Fig. \ref{fig:density}, we show samples of correlation functions $\rho_m(x)$ of different orders $m$ calculated for the parameters $ng=25$, $N=500$, and $K=50,\, 125,\, 250$. Densities $\left|\phi_{\rm MF}(x)\right|^2$ of the solitonic solution of Eq.~\eqref{eq:NLSE} are shifted so that their notch positions overlap with that of $\rho_{150}(x)$ for direct comparison. We observed that the notch position of $\rho_{m}(x)$ is determined early i.e. for low order correlation functions and stabilizes as $m$ increases, with slight fluctuations dependent on the random particle position draws. Even highly disruptive particle measurements caused by unlikely draws, as that exemplified by $\rho_{50}(x)$ for $K=100$ in Fig. \ref{fig:density}, do not prevent the formation of a dark soliton.
It is important to mention that every curve presented in Fig. \ref{fig:density} is a result obtained for a single simulation. In all simulations we have made, the high order correlation functions always resemble the density of a MF soliton.
Note, that our result corresponds to the short healing length $\xi=0.2L$. We observe that the MF soliton emerges from the correlation function as $m$ increases. We also notice that obtaining a very good agreement between the density emerging from the many-body calculation and the MF soliton requires calculation of high order correlation function. 
The agreement between the MF approach and the Ansatz encourages us to investigate the spatial phase, which is peculiar for dark solitons. A phase ${\rm arg}\left\{\phi_{\rm MF}(x) \right\}$ of the soliton changes quickly within the density notch, but remains linear far from it (see  the thick gray line in the right panel of Fig.~\ref{fig:density}). Therefore, when the system is in the solitonic state, the majority of atoms moves along the circle with a constant velocity, apart from the place of rarefaction where particles move quickly in the opposite direction. We extract the phase of the many-body wave function using
		\begin{eqnarray}
	&& {\rm g}_1(x) \propto \label{eg:g1} \\ &&\langle \oppsi^{\dagger}(x_1)...\oppsi^{\dagger}(x_{m-1})\oppsi^{\dagger}(x) \oppsi(0) \oppsi(x_{m-1})...\oppsi(x_1) \rangle,
	\nonumber
	\end{eqnarray}
 evaluated for the Ansatz~\eqref{eq:smeared-sol} (for details of our calculations see Appendix \ref{appendix:computation-state}).
	As shown in the right panel of Fig.~\ref{fig:density}, the phase of $\text{g}_1$ converges to $\arg \left\{\phi_{\rm MF}(x)\right\}$ for increasing $m$.  The phase of MF soliton is shifted by the same amount as the corresponding density $\left|\phi_{\rm MF}(x)\right|^2$ in the left panel.

	The above results prove that, indeed, the MF solitons emerge  in the high order correlation function evaluated for the Ansatz~\eqref{eq:smeared-sol}  in the true MF regime with $\xi$ significantly smaller than $L$.

	On the other hand, in Refs~\cite{Sato2012, Sato2016, Kaminishi2019} the MF solitons were constructed from the many-body eigenstates of the LL Hamiltonian~\eqref{eq:LLham} in a completely different way, as explain in the next section.
	

\section{Dark solitons as superpositions of yrast states}\label{sec:superposition}

 In the previous section, we have shown using our Ansatz that the MF  solitons emerge in high order correlation functions.   An opposite direction was taken in Refs.~\cite{Sato2012, Sato2016, Kaminishi2019, Brand2019} where the dark solitons are constructed as a specific superposition of yrast states. Namely the MF product state is expressed as:
\begin{equation}
\ket{\phi_{\rm MF}}^{\otimes N} \approx \sum_{K'}a_{K'}\ket{K'}_N
	\label{eq:japaneseidea}
\end{equation}
where $a_{K'}$ are expansion coefficients ( drawn from a chosen distribution) and $\ket{K'}_N$ denotes a yrast state of the system with $N$ particles and the total momentum $\frac{2\pi\hbar}{L}K'$. A comprehensive discussion of different $a_{K'}$ choices can be found in Refs.~\cite{Sato2012, Sato2016, Kaminishi2019,Brand2019}.

An interesting question arises whether these two approaches of linking the yrast states of the LL model with the dark solitons from the NLSE complement each other or are completely unrelated. We shall answer this question by appealing to the definition of the high order correlation function $\rho_m$~\eqref{eq:mordercor} and using the Ansatz~\eqref{eq:smeared-sol} for the yrast state.

Calculation of any correlation function by means of the second quantization requires the sequential action of the annihilation field operators at some points in space. One can say that such procedure conditions a system's wave function. Physically, it corresponds to an instantaneous destructive measurement of certain particle positions. Thus, we introduce a conditional wave function $\ket{\tilde{\psi}^m}$  of the system in a state $\ket{\psi}$ after measuring positions of $m$ particles given by
\begin{equation}
\ket{\tilde{\psi}^m} \propto \oppsi(x_m)  \oppsi(x_{m-1}) \ldots \oppsi(x_1) \ket{\psi}.
\label{eq:psi-k}
\end{equation}
To maximize the reliability of such a measurement in any theoretical considerations it has to be performed according to a multivariate probability distribution determined by the wave function for a given state. Therefore, each position $x_i$ from Eq.~\eqref{eq:psi-k} should be taken from the particular PDF $\rho_i$ defined in Eq.~\eqref{eq:mordercor}.

The average density in the conditional state \eqref{eq:psi-k} $\rho(x):=\meanv{\oppsi^{\dagger}(x) \oppsi(x)}$ is equal to the $\bb{m+1}$-th order correlation function $\rho_{m+1}$ \eqref{eq:mordercor}
 studied in the previous section.
Therefore, to bridge the different views on the correspondence between the MF solitons and yrast states, one has to verify whether 
the conditional wave function~\eqref{eq:psi-k} for $\ket{\psi}$ being a yrast state can be  represented as a  wave packet of yrast states, each with $N-m$ atoms and different total momentum. As calculations with the help of the exact many-body states would be limited to a small number of atoms, we again refer to the family of Ansatzes~\eqref{eq:smeared-sol} as an approximation for the yrast states with different total momenta $\frac{2\pi\hbar}{L} K'$.

For the Ansatz~\eqref{eq:smeared-sol} one can easily find the conditional state \footnote{For details of our calculations see Appendix \ref{appendix:computation-state}.}
\begin{eqnarray}		
&\ket{\tilde{\psi}^m_{\rm Ansatz}} \propto\oppsi(x_m)\oppsi(x_{m-1}) \ldots \oppsi(x_1) \ket{\psi_{\rm Ansatz}} \propto  \nonumber\\ & \int_{0}^{L} dy\,e^{i \frac{2\pi}{L} K y}\label{eq:psitilde-Ansatz-k} \bb{\prod_{j=1}^m \phi_{\rm MF}(x_{j}-y)} e^{-i \hat{P} y/\hbar} \ket{\phi_{\rm MF}}^{\otimes (N-m)}. 
\end{eqnarray}
Due to the factors  $\phi_{\rm MF}(x_j-y)$, the solitons centered close to the positions $x_{j}$, where a measurement occurred, enter the conditional state with lower weights, as compared to solitons with a density dip far from $x_{j}$.
The conditional wave function $\ket{\tilde{\psi}^m_{\rm Ansatz}}$ is no longer an eigenstate of LL system with $N-m$ particles as it is not translationally invariant. However, we can always decompose this state into the set of eigenstates of LL model for $N-m$ particles in the following way:
\begin{equation}
\ket{\tilde{\psi}^m_{\rm Ansatz}}=\sum_{K'}a_{K'}\ket{K'}_{N-m}+\sum_{j}b_j\ket{\psi_j}_{N-m}
\label{eq:decomposition}
\end{equation}
where $\ket{\psi_j}_{N-m}$ is an eigenstate of the system, which is not the yrast state, and  $\sum_{K^{\prime}}|a_{K^{\prime}}|^2+\sum_{j}|b_j|^2=1$. 
The question is whether the conditional state $\ket{\tilde{\psi}^m_{\rm Ansatz}}$ remains in the subspace of the yrast states in the form of a wave packet.
To answer the question we calculate the overlap between the Ansatz \eqref{eq:smeared-sol} and conditional state \eqref{eq:psitilde-Ansatz-k} finding the weight of the yrast subspace given by $\sum_{K^{\prime}}|a_{K^{\prime}}|^2$ and the $a_K'$ distribution \footnote{
For details of our computations see Appendix \ref{appendix:overlap-after-measurmenet}}.

\begin{center}
		\begin{figure}[h!]
            \includegraphics[]{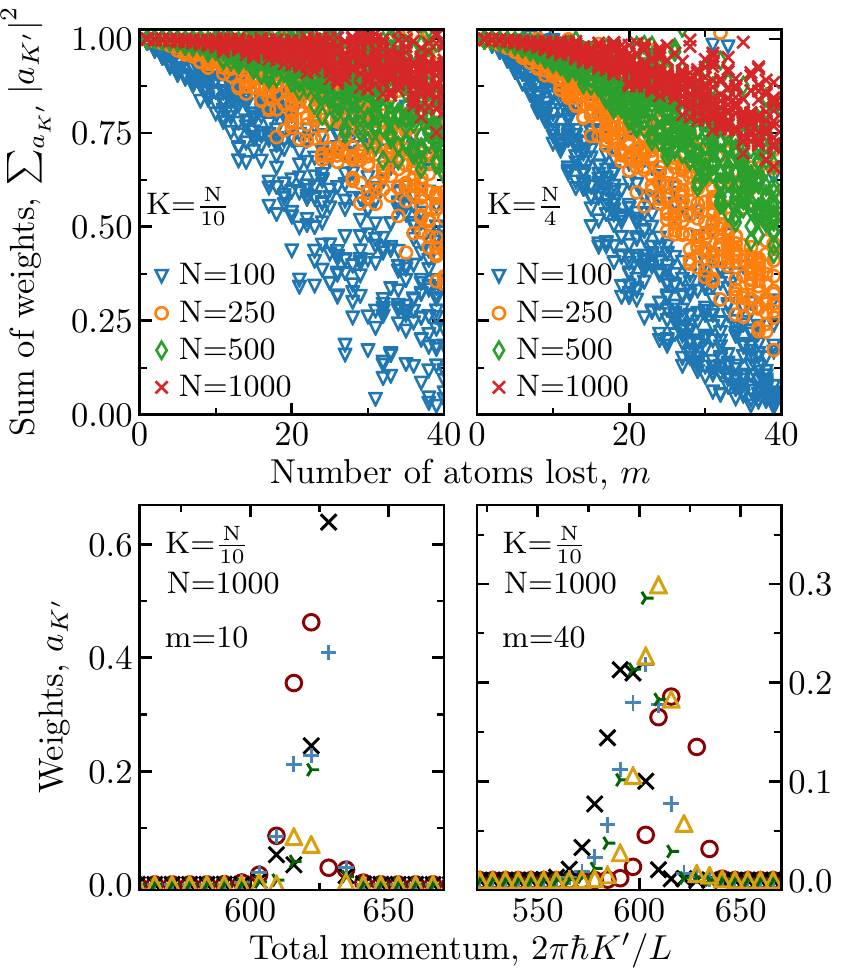}
			\caption{(color online) Top: Sum of weights of the yrast subspace $\sum_{K^{\prime}} \left|a_{K^{\prime}}\right|^2$ as a function of the number of atoms lost $m$.
			Each value for a given $m$ corresponds to a different stochastic sequence of particle positions measured $\{x_i\}$ obtained for $N=100\,\text{(triangles)},$ $250\,\text{(circles)}$, $500\,\text{(rhombus)},$ $1000$ (crosses), with $K$ set to $N/10$ in the left panel and $N/4$ in the right one. Bottom: Five representative distributions of weights $a_{K^{\prime}}$ as functions of the total momentum for $m=10$ (left) and $m=40$ (right). 
			Each symbol corresponds to a different set $\{x_i\}$. Parameters $N$ and $K$ are set to $1000$ and $N/10$, respectively. The mean-field parameter $ng$ is equal to $25$ for every graph.
			\label{fig:weights}}
		\end{figure}
	\end{center}

In the top panel of Fig.~\ref{fig:weights}, we present sum of weights $\sum_{K^{\prime}}|a_{K^{\prime}}|^2$ of the yrast subspace as a function of a measured number of atoms $m$ for $ng=25$, $N=100,\, 250,\, 500,\ 1000$ and a fixed total momentum $K=N/10$ (left) and $K=N/4$ (right). We observe that the greater the number of atoms $N$ the closer to one the weight of the yrast subspace for a given value of $m$ is. It means that, indeed, the conditional wave function, which reveals the dark soliton, is approximated by a superposition of the yrast states. In the bottom panel of Fig.~\ref{fig:weights}, we plot five representative distributions of $a_{K^{\prime}}$ as functions of total momentum $\frac{2\pi\hbar}{L}K^{\prime}$ for $ng=25$, $N=1000$, $K=N/10$, $m=10$ (left) and $m=40$ (right). 
The resulting distributions differ from shot-to-shot but they give the same single-particle density. Note that this agrees with the existing literature where different distribution models were considered.

Our efforts in bridging the dark solitons and the yrast states also result in the unification of the previous attempts done in the literature~\cite{Sato2012, Sato2016, Kaminishi2019, Brand2019}. In this section we have shown that the dark solitons hosted in the non-ideal gas of $N$ atoms and revealed by partial measurements can be almost exactly expressed as a wave packet of the yrast states of a gas of $N-m$ atoms.
	
\section{Conclusions}

We studied the correspondence between the yrast states of the Lieb-Liniger Hamiltonian and the
mean-field solitons from the non-linear Schrödinger equation. To this end we proposed a simple construction for the yrast state~\eqref{eq:smeared-solv2} based on mean-field product states with appropriate phase factors. Using this Ansatz we were able to unify previous literature results and observations~\cite{syrwid2015, Sato2012, Sato2016, Kaminishi2019, Brand2019} about the subject at hand. 

The conditional wave function, which results from annihilation of $m$ particles in the Ansatz state at random positions, reveals the ultimate utility of our approach. The single-particle density evaluated in the conditional wave function is the $m$-th order correlation function resembling the mean-field soliton, as discussed in Ref.~\cite{syrwid2015}. Moreover, the conditional wave function is found to be a wave packet of yrast states of $N-m$ atom system with different total momenta, as analysed in Refs.~\cite{Sato2012, Sato2016, Kaminishi2019, Brand2019}. As can be readily seen, our proposal complements various preceding studies reproducing their results with a singular construction and thus tying them into a single picture.

The "measurements" needed to break the translational symmetry could be realized spontaneously, due to particle losses which are inevitable in the ultracold gases. We plan to study this in detail, using many-body methods. 
Another remaining question concerns dynamical stability of the conditional wave function. 
In all previous works, the emerging solitonic profiles were, unlike the mean-field solitons, blurred during evolution \cite{syrwid2016, Sato2012, Kaminishi2019, Brand2019}. It was argued that the time of blurring should increase to infinity in the thermodynamic limit \cite{syrwid2016, Sato2016}. However, this hypothesis has to be verified. This can be done again employing our Ansatz.

	\begin{acknowledgments}
	We acknowledge fruitful discussions with K.	Rz\k a\.zewski, M. Gajda and K. Sacha.
	In particular we thank to M. Gajda for convicing us, that the results concerning the unification of different views on relations between yrast states and MF solitons are the most interesting aspects of our work.
	This work was supported by the (Polish) National Science Center Grants 2016/21/N/ST2/03432 (R.O.),
	2015/19/B/ST2/02820 (W. G\'orecki), 2014/13/D/ST2/01883 (W. Golletz and K.P.) and  2018/31/B/ST2/00349 (W. Golletz).
	\end{acknowledgments}

	\appendix
	\section{Mean-field solitons\label{appendix:mean-field}}
	\subsection{Mean-field gray solitons}
The solitonic solution $\phi_{\rm MF}(x)$ of NLSE~\eqref{eq:NLSE} was discussed several times in literature \cite{zakharov73, Carr2000, Sato2016}. Here we briefly present the final formulas, in the form which was used to produce results of this paper.

We followed a procedure described in \cite{Sato2016}. Solitonic solution of NLSE is a running wave, 
$\phi_{\rm MF}(x,\;t) = \phi_{\rm MF}(x - vt)$ with speed $v$. In what follows we will omit the time dependence and give separately a solitonic density:
\begin{equation}
    \rho(x) = |\phi_{\rm MF}(x)|^2,
\end{equation}
and its phase
\begin{equation}
    \varphi(x) = {\rm Arg}\left\{\phi_{\rm MF}(x)\right\},
\end{equation}
and its speed $v$.
The approach presented in \cite{Sato2016} was devoted to the case of gray solitons, i.e. when $\varphi(x)$ is continuous and $\rho(x)$ is always larger than $0$.
The density and the phase and the velocity of the dark soliton obeying NLSE, in terms of four parameters denoted with $a_1$, $a_2$, $a_3$ and $k$, are given by:
\begin{eqnarray}
    \rho(x) &=& \bb{a_1 + \bb{a_2-a_3}\,{\rm sn}^2 \bb{\sqrt{g} \sqrt{a_3-a_1}x},
    \,k}/N
    \nonumber\\
    \varphi(x) &=& \frac{v}{2} x + \frac{\sqrt{a_2a_3\,\Pi \bb{1-\frac{a_2}{a_1}, {\rm am}\bb{\sqrt{g}\sqrt{a_3-a_1}x},\,k}}}{\sqrt{a_1 \bb{a_3-a_1}}}\nonumber\\
    v & = & \frac{4\sqrt{a_2 a_3}}{L\sqrt{a_1 \bb{a_3-a_1}}}\Pi\bb{1-a_2/a_1,\,k}.
    \label{eq:appendix:densityMF}
\end{eqnarray}
where ${\rm sn} (u,\,k)$ is the Jacobi elliptic function, $\Pi$
is the incomplete elliptic integral of the third kind, with the modulus of Jacobi's elliptic function $k$ and ${\rm am}$ is the Jacobi amplitude. 
Parameter $a_1$ has simple physical interpretation --- it is the minimal density in the solitonic solution. 
Periodic boundary conditions for the phase and density and normalization condition $\int |\phi_{\rm MF}|^2 = 1$  
lead to the following relations between parameters $a_1$, $a_2$, $a_3$ and the elliptic modulus $k$:
\begin{eqnarray}
    a_1 &=& n + \frac{4 K(k) \bb{E(k) - K(k)}}{L^2 g}\label{eq:appendix:a1}\\ 
    a_2 &=& n + \frac{4 K(k) \bb{E(k) - (1-k^2)K(k)}}{L^2 g}\label{eq:appendix:a2}\\
    a_3 &=& n + \frac{4 K(k) E(k) }{L^2 g} \label{eq:appendix:a3},
\end{eqnarray}
where $K(k)$ and $E(k)$ are the elliptic integrals of the first and the second kind, respectively.

To compute phase and density of a soliton that has a desired minimum of the density $a_1$, we  first solve numerically Eq. \eqref{eq:appendix:a1} for the elliptic modulus $k$, and then we use Eqs. \eqref{eq:appendix:a2} and \eqref{eq:appendix:a3} to find the remaining parameters $a_2$ and $a_3$. Having determined $a_1$, $a_2$, $a_3$ and $k$ we can compute the soliton wave function $\phi_{\rm MF}(x)$ at any position $x$, using Eqs.  \eqref{eq:appendix:densityMF} and relation $\phi_{\rm MF}(x) = \sqrt{\rho(x)}e^{i\varphi(x)}$. The average momentum $\langle \hat{p}\rangle :=\frac{\hbar}{i}\int\,{\rm d}x\, \phi_{\rm MF}^*(x) \partial_x\phi_{\rm MF}^*(x)
$ is then computed numerically.

To find the MF soliton with target momentum $p_{\rm target} = 2\pi \hbar K /(N L)$, we repeat the steps described above varying $a_1$, until the numerically determined momentum matches $p_{\rm target}$. After a few bisection steps with respect to $a_1$ the target MF soliton is found -- we stop bisection when relative discrepancy between numerically computed momentum and the target momentum is below $10^{-6}$.

\subsection{Mean-field black solitons}

The formulas presented in the previous section are derived under the assumption that the density $\rho(x)$ given in Eq. \eqref{eq:appendix:densityMF} is always grater than $0$. Therefore, they can not be used in the case of a black soliton.
In fact, already for a gray but very deep solitons, the equation \eqref{eq:appendix:a1} becomes very demanding, as discussed in \cite{Sato2016} (see for instance Table 4 in \cite{Sato2016} for the values of the parameter $k$).

On the other hand, one can use the properties of the black soliton to quickly find it numerically with other method. 
The trick is to compute the lowest energy state of NLSE~\eqref{eq:NLSE}, but in the space of functions with a given phase. We have learned this trick from Tomasz Karpiuk~\footnote{Private communication.}, and used it successfully in \cite{dipoloweSolitony2015}.
Precisely, we look for solutions with the phase:
\begin{equation}
    \varphi_{\rm black}(x) := \pi \bb{{\rm sgn}\bb{x - L/2} - x/L},
    \label{eq:appendix:argblack}
\end{equation}
where the signum function ${\rm sgn}$ is equal to $1$ for positive arguments and $0$ for negative ones. The signum function introduces a discontinuity in the phase, a $\pi$ jump, which is the characteristic feature of a black soliton.

The minimal energy state in the space of functions with phase $\varphi_{\rm black}(x)$ is found with the split-step imaginary time evolution, implemented as follows:
\begin{itemize}
    \item[1)] We start with an arbitrary function $\phi(x)$.
    \item[2)] We compute $\tilde{\phi}(x)$  according to the split-step formula:
    \begin{equation}
        \tilde{\phi}(x) := e^{ -\hat{T}\, \delta t} e^{ -\hat{V}  \, \delta t} \phi(x), 
        \label{eq:appendix:bs_1}
    \end{equation}
    where $\hat{V} = g |\phi(x)|^2$ and 
    $\hat{T} = -\frac{ \hbar^2 }{2m} \partial_x^2$.
    To act with the operator $e^{ -\hat{T}\, \delta t}$ we apply  Fourier transform $\mathcal{F}$ and its inverse $ \mathcal{F}^{-1}$:
    \begin{eqnarray}
        \tilde{\phi}(x) &=& \mathcal{F}^{-1}\left[\mathcal{F} e^{ -\hat{T}\, \delta t} e^{ -\hat{V}  \, \delta t} \phi(x)\right]= \nonumber\\
        &=&\mathcal{F}^{-1}\left\{\mathcal{F} \left[e^{ -\hat{T}\, \delta t}\right] \mathcal{F} \left[ e^{ -\hat{V}  \, \delta t} \phi(x)\right]\right\},
    \end{eqnarray}
    to replace the cumbersome operator  $e^{ (\hbar^2 \delta t/ 2 m ) \partial^2_x }$ with its Fourier representation.
    
    \item[3)] We normalize the output of the previous step: 
    \begin{equation}
        \tilde{\tilde{\phi}}(x) := \frac{\tilde{\phi}(x)}{\int {\rm d}x\,|\tilde{\phi}(x)|^2}.
    \end{equation}
    \item[4)] We define a new function $\phi(x)$ as $\tilde{\tilde{\phi}}(x)$ but with the phase "overwritten" with $\varphi_{\rm black}$: 
    \begin{equation}
        \phi(x) = |\tilde{\tilde{\phi}}(x)|\,e^{ i \varphi_{\rm black}(x)}.
        \label{eq:appendix:bs_4}
    \end{equation}
\end{itemize}
We repeat steps (2)--(4) until the energy of $\phi(x)$ converges.

\begin{figure}[]
    \centering
			\includegraphics[]{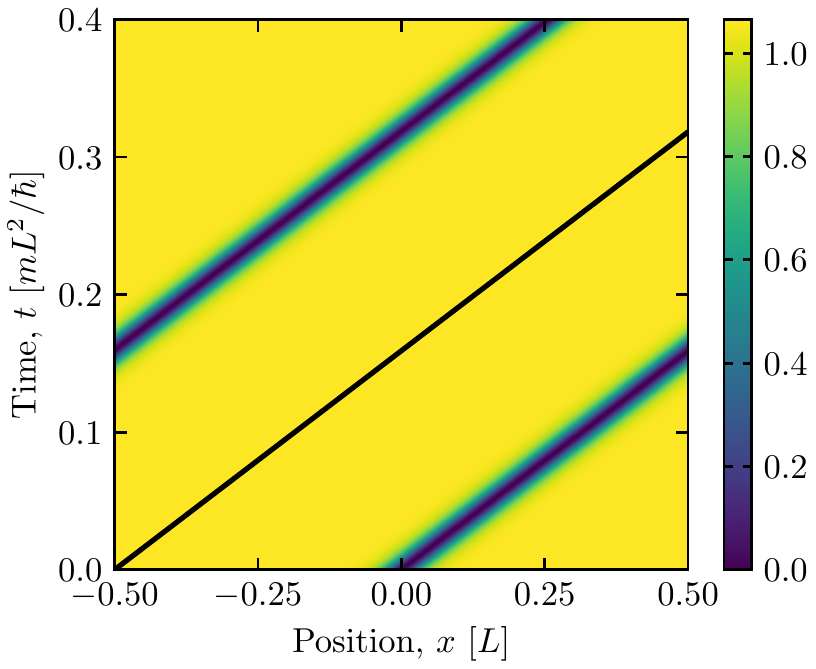}
			\caption{(color online) Evolution of the density $\left| \phi(x, t)\right|^2$ governed by non-linear Schr\"{o}dinger equation~\eqref{eq:NLSE}. The mean-field black soliton $\phi(x, t=0)$ with density dip at $x=0$, used as an initial state, was found numerically according to steps \eqref{eq:appendix:bs_1}--\eqref{eq:appendix:bs_4}. The black line is the reference line - a trajectory of a point moving with the speed $v_{\rm black} = \hbar \pi/ m L$.}
    \label{fig:appendix:xt}
\end{figure}

The procedure described above wasn't proven to give the exact result,  although it may be rooted in the relations between an yrast state and a ground state solution for the interacting bosons placed in a one-dimensional hard-wall box potential \cite{Reichert2019}, found by Gaudin \cite{Gaudin1971}. We further verify numerically if the final state $\phi(x)$ is indeed the solution of the NLSE~ \eqref{eq:NLSE}. We use $\phi(x)$ as an initial state  $\phi(x,\,t=0)$ for the Eq.~\eqref{eq:NLSE} and check whether $\left|\phi(x, t>0)\right|^2$ preserves its shape during evolution and whether
the density dip moves with the expected speed. Example of such verification is presented in Fig. \ref{fig:appendix:xt}. Additionally we compare $\phi(x)$ with the series of the deepest solitonic solution we were able to find with the methods for gray solitons described in  the previous section to check if they converge to the black soliton found with the method describe in this section. 

	\begin{widetext} 
\section{The Ansatz\label{appendix:ansatz}}

	In the main text we use the following Ansatz for a yrast state $\ket{K}$:
	\begin{equation}
	\psi_{\rm Ansatz}(x_1,\ldots, x_N) = \mathcal{N}\int_0^L\,dy\,e^{i \frac{2\pi}{L}K y} \prod_{j=1}^N \phi_{\rm MF} (x_j-y),
	\label{eq:appendix:smeared-sol}
	\end{equation}
where $\mathcal{N}$ is a normalization factor, and $\phi_{\rm MF} (x)$ is the solitonic solution of the NLSE~\eqref{eq:NLSE} which has the average momentum equal to $\frac{2\pi \hbar}{L}K/N$, 
\begin{equation}
\meanv{\hat{p}} :=-i \hbar\int_0^L dx \,\phi_{\rm MF}^* (x)\,\partial_x\phi_{\rm MF} (x) = \frac{2\pi \hbar}{L}K / N.
\end{equation} 

\noindent The normalization factor $\mathcal{N}$  from Eq.~\eqref{eq:appendix:smeared-sol} is evaluated from the normalization condition:
\begin{eqnarray}
1 = \left\langle \psi_{\rm Ansatz} |\psi_{\rm Ansatz}\right\rangle & = & \mathcal{N}^2\int_0^L dy'\,\int_0^L dy\, e^{i \frac{2\pi}{L}K (y - y')}  \nonumber
\bb{ \bra{\phi_{\rm MF}}^{\otimes N} e^{-i \hat{P} (y-y')/\hbar} \ket{\phi_{\rm MF}}^{\otimes N} }\\
& = &\mathcal{N}^2\int_0^L dy'\,\int_0^L dy\, e^{i \frac{2\pi}{L}K (y - y')} 
\bb{ \int_0^L dx \;\phi_{\rm MF}^* (x) \, \phi_{\rm MF}(x-y+y') }^N,
\end{eqnarray}
where the overlap $\int_0^L dx\, \phi_{\rm MF}^* (x) \,\phi_{\rm MF}(x-y+y') $ and integrals over $y$ and $y'$ are evaluated numerically. 

\subsection{The Ansatz as an eigenstate of the total momentum operator \label{appendix:ansatz-momentum}}

Let us start with a comment that  MF product state
\begin{equation}
    \prod_{j=1}^N \phi_{\rm MF} (x_j)
    \label{eq:appendix_MF}
\end{equation}
does not have well defined momentum, i.e. it is a wave packet of eigenstates with different momenta. In contrast the Ansatz  ~\eqref{eq:appendix:smeared-sol} is an eigenstate of the momentum operator.
To prove that we begin by showing that translation of all particles by an 
arbitrary shift $\Delta x$ is equivalent to multiplication by a global phase factor:
\begin{eqnarray}
	\psi_{\rm Ansatz}(x_1+\Delta x,&\ldots&, x_N+\Delta x) = \mathcal{N}\int_0^L\,dy\,e^{i \frac{2\pi}{L}K y} \prod_{i=1}^N \phi_{\rm MF}(x_i+\Delta x-y)\nonumber\\
	&=& \mathcal{N}\int_0^L\,dy\,e^{i \frac{2\pi}{L}K y} \prod_{i=1}^N \phi_{\rm MF}(x_i-(y-\Delta x))	\overset{y'=y-\Delta x}{=}\mathcal{N}\int_{-\Delta x}^{L-\Delta x}\,dy'\,e^{i \frac{2\pi}{L}K (y'+\Delta x)} \prod_{i=1}^N \phi_{\rm MF}(x_i-y')\nonumber\\
		&=&e^{i\frac{2\pi}{L}K\Delta x}\mathcal{N}\int_{0}^{L}\,dy'\,e^{i \frac{2\pi}{L}K y'} \prod_{i=1}^N \phi_{\rm MF}(x_i-y') =e^{i\frac{2\pi}{L}K\Delta x}\psi_{\rm Ansatz}(x_1,\ldots, x_N).
\end{eqnarray}
In the second last equality we have shifted the integration limits with no impact on its value due to the periodicity of integrated function.

Using the relation above, one can explicitly check that $\psi_{\rm Ansatz}$ is indeed an eigenstate of $\hat{P}$ with corresponding eigenvalue $\frac{2\pi\hbar}{L}K$:
\begin{eqnarray}
	\hat{P}\psi_{\rm Ansatz}(x_1,\ldots,x_N)&=& \hat{P}\psi_{\rm Ansatz}(x_1+\Delta x,\ldots,x_N+\Delta x)\Big|_{\Delta x=0}= -i\hbar\sum_{j=1}^N\partial_{x_j}\psi_{\rm Ansatz}(x_1+\Delta x,\ldots,x_N+\Delta x)\Big|_{\Delta x=0}\nonumber\\
	&=&-i\hbar\partial_{\Delta x}\psi_{\rm Ansatz}(x_1+\Delta x,\ldots,x_N+\Delta x)\Big|_{\Delta x=0}=-i\hbar\partial_{\Delta x}\left(e^{i\frac{2\pi}{L}K\Delta x}\psi_{\rm Ansatz}(x_1,\ldots, x_N)\right)\Big|_{\Delta x=0}\nonumber\\
    &=&\frac{2\pi\hbar}{L}K\psi_{\rm Ansatz}(x_1,\ldots,x_N).
\end{eqnarray}
\noindent Note that the Ansatz is an eigenstate of the total momentum operator, irrespectively of the choice of the orbital $\phi(x)$. As discussed in the main text, only with an appropriate choice of orbital does the Ansatz become good approximation of the yrast state.

\subsection{Comparisons between the mean-field product state and Ansatz in the limit $g\to 0$
\label{appendix:ideal-gas}}

As stated before, the MF product state~\eqref{eq:appendix_MF} is not a state with well-defined momentum and therefore it cannot be a good approximation of the exact yrast state. That is why we have decided to consider a properly weighted, by a phase factor, superposition of MF solitons.
To get some intuition on how these two are related, it is convenient to discuss their properties in the limit $g\to 0$. Here we discuss the case with mean total momentum $\frac{2\pi\hbar}{L}\frac{N}{2}$, as for this one the analytical formulas are the simplest. 
The MF orbital is:
\begin{equation}
    \phi^{g\to 0}_{\rm MF}(x)=\frac{1+e^{+i\frac{2\pi}{L}x}}{\sqrt{2L}},
\end{equation}
which we denote symbolically as $\ket{\phi^{g\to 0}_{\rm MF}}=\frac{1}{\sqrt{2}}(\ket{0}+\ket{2\pi\hbar/L})$. Corresponding $N$-particle product state is a superposition of the yrast states with coefficients given by square roots of the binomial distribution coefficients:
\begin{equation}
   \ket{\phi^{g\to 0}_{\rm MF}}^{\otimes N}=\left(\frac{1}{\sqrt{2}}(\ket{0}+\ket{2\pi\hbar/L})\right)^{\otimes N}\\
   =\frac{1}{\sqrt{2}^N}\sum_{k=0}^N\sqrt{\binom{N}{k}}\ket{n_0:N-k,n_{2\pi\hbar/L}:k},
\end{equation}
where $\ket{n_0:N-k,n_{2\pi\hbar/L}:k}$ denote the Fock state with $N-k$ atoms in orbital with momentum $0$ and $k$ atoms in obital with momentum $2\pi\hbar/L$. Similar analysis may also be done for any gray soliton \cite{Kaminishi2019}. The expectation value of kinetic energy is, as expected, the same as for the exact yrast state $\ket{n_0:N/2,n_{2\pi\hbar/L}:N/2}$:
\begin{equation}
    \braket{\phi^{g\to 0}_{\rm MF}}{{}^{\otimes N}\hat E_{\rm kin}|\phi^{g\to 0}_{\rm MF}}^{\otimes N}=\frac{1}{2^N}\sum_{k=0}^N\binom{N}{k}\frac{k(2\pi\hbar)^2}{2mL^2}=\frac{2\hbar^2\pi^2}{mL^2}\frac{N}{2}.
\end{equation}
However, for the mean value of interaction energy situation turns out to be slightly more complicated. We get:
\begin{eqnarray}
     \braket{\phi^{g\to 0}_{\rm MF}}{{}^{\otimes N}\hat E_{\rm int}|\phi^{g\to 0}_{\rm MF}}^{\otimes N}&&=
     \frac{1}{\sqrt{2}^N}\sum_{k=0}^N\binom{N}{k}\bra{n_0:N-k,n_{2\pi\hbar/L}:k}\hat E_{\rm int}\ket{n_0:N-k,n_{2\pi\hbar/L}:k} \nonumber\\
     &&=\frac{1}{\sqrt{2}^N}\sum_{k=0}^N\binom{N}{k} \frac{g}{2L}(N(N-1)+2Nk-2k^2)=
     \frac{gN}{4L}(3N-3),
\end{eqnarray}
where in the first step we have used the fact than interaction energy operator $\hat E_{\rm int}$ preserve the total momentum of the system and therefore $\forall_{k\neq k'}\bra{n_0:N-k,n_{2\pi\hbar/L}:k}\hat E_{int}\ket{n_0:N-k',n_{2\pi\hbar/L}:k'}=0$.
On the other hand, interaction energy of the yrast state $\ket{n_0:N/2,n_{2\pi\hbar/L}:N/2}$ is:
\begin{equation}
     \braket{n_0:N/2,n_{2\pi\hbar/L}:N/2}{\hat E_{\rm int}|n_0:N/2,n_{2\pi\hbar/L}:N/2}=\frac{gN}{4L}(3N-2).
\end{equation}
We see, that the expectation value of the mean-filed product state's energy is smaller than the energy of the yrast state. It must be so, as the formula for interaction energy for the Fock state $\ket{n_0:N-k,n_{2\pi\hbar/L}:k}$: $\frac{g}{2L}(N(N-1)+2Nk-2k^2)$ takes its maximum in $K=N/2$, therefore increasing the impact of other Focks in MF product state may only decrease the energy.

We want to stress again, that this result does not lead to contradiction with the definition of the yrast state (i.e. the state with the lowest energy for given total momentum), as the MF product state is not an eigenstate of total momentum operator.

\subsection{Conditional states, single-particle densities and function $g_1$ after measurement of particle positions  \label{appendix:computation-state}}

The measurement of the particle positions is expressed by an action of the field operator $\hat{\Psi}(x)$ on the Ansatz.
To some extent it can be evaluated analytically. If the particle has been measured at random position $x_1$, then the conditional wave function $\ket{\tilde{\psi}^1_{\rm Ansatz}}$ is given by:
\begin{eqnarray}
\ket{\tilde{\psi}^1_{\rm Ansatz}} \propto \oppsi(x_1)\ket{\psi_{\rm Ansatz}}  &=&  \mathcal{N}\int_0^L\,dy\,e^{i \frac{2\pi}{L}K y} \oppsi(x_1) e^{-i \hat{P} y/\hbar} \ket{\phi_{\rm MF}}^{\otimes N} \nonumber \\ 
& =&\mathcal{N}\int_0^L\,dy\,e^{i \frac{2\pi}{L}K y} \,\oppsi(x_1)\,  \ket{\phi_{\rm MF}(x-y)}^{\otimes N} \nonumber \\
& =&\mathcal{N}\int_0^L\,dy\,e^{i \frac{2\pi}{L}K y} \bb{\sqrt{N} \phi_{\rm MF}(x_1-y) \ket{\phi_{\rm MF}(x-y)}^{\otimes (N-1)}},
\label{eq:appendix:psi-1}
\end{eqnarray}
where we used the fact that $\oppsi(x)\ket{f}^{\otimes N}  = f(x) \ket{f}^{\otimes (N-1)}$,
and  introduced the proportionality symbol $\propto$ because the state $\oppsi(x_1)\ket{\psi_{\rm Ansatz}}$ is not normalized.

By repetitive action of the field operator we can write down a conditional state after $m$ subsequent measurements which occurred at random positions $x_1,\,x_2\,\ldots,\,x_m$:
\begin{equation}
\ket{\tilde{\psi}^m_{\rm Ansatz}} \propto \bb{\prod_{j=1}^m\oppsi(x_j)}\ket{\psi_{\rm Ansatz}}  \propto \int_0^L\,dy\,e^{i \frac{2\pi}{L}K y} \bb{\prod_{j=1}^m \phi_{\rm MF}(x_j-y)} \ket{\phi_{\rm MF}(x-y)}^{\otimes (N-m)}.
\label{eq:appendix:psik}
\end{equation}

Given the conditional wave function, we write down its single particle density which is equal to ($m+1$)-th order correlation functions used in the main text:
\begin{eqnarray}
\rho_{m+1}(x)&:=&\meanv{\oppsi^{\dagger}(x)\oppsi(x)} = \left\|\oppsi(x)\ket{\tilde{\psi}^m_{\rm Ansatz}} \right\|^2  \propto   \int_0^L\,dy\int_0^L\,dy'\left[ \,e^{i \frac{2\pi}{L}K (y-y')} \bb{\prod_{j=1}^m \phi^*_{\rm MF}(x_j-y')\phi_{\rm MF}(x_j-y)} \right. \nonumber\\
 &  & \left. \quad \times \phi^*_{\rm MF}(x-y')\phi_{\rm MF}(x-y)
 \left(\int_0^L dz \phi^*_{\rm MF}(z-y')\phi_{\rm MF}(z-y)\right)^{(N-m-1)}
 \right],\label{appendix:rho_m}\\
\text{g}_1(x) &:=& \meanv{\oppsi^{\dagger}(x)\oppsi(0)}    \propto  \int_0^L\,dy\int_0^L\,dy'\left[ \,e^{i \frac{2\pi}{L}K (y-y')} \bb{\prod_{j=1}^m \phi^*_{\rm MF}(x_j-y')\phi_{\rm MF}(x_j-y)} \right. \nonumber\\
&  & \left. \quad \times \phi^*_{\rm MF}(x-y')\phi_{\rm MF}(-y) 
\left(\int_0^L dz \phi^*_{\rm MF}(z-y')\phi_{\rm MF}(z-y)\right)^{(N-m-1)}
\right].
\end{eqnarray}


\subsection{Interaction and kinetic energy \label{appendix:energy}}

Here we discuss our procedure of computing the interaction energy:
\begin{equation}
E_{\rm int} = \frac{g}{2}\,\int_0^{L}\,{\rm d}x\,\left\langle \psi_{\rm Ansatz} 
| \hat{\Psi}^{\dagger} (x)\hat{\Psi}^{\dagger} (x)\hat{\Psi} (x)\hat{\Psi} (x) |  \psi_{\rm Ansatz}\right\rangle = \frac{g}{2}\,\int_0^{L}\,{\rm d}x\,\left\| \hat{\Psi}^2 (x) |  \psi_{\rm Ansatz}\rangle \right\|^2. 
\label{eq:appendix:ansatz-eint}
\end{equation}
where the square of the norm is in fact a double integral:
\begin{eqnarray}
\label{eq:appendix:ansatz-eint3}
&&\left\| \hat{\Psi}^2 (x) |  \psi_{\rm Ansatz}\rangle \right\|^2=\\& &\mathcal{N}^2 N(N-1)
 \int_0^L\,dy\int_0^L\,dy' \,e^{i \frac{2\pi}{L}K (y-y')} \left(\phi^*_{\rm MF}(x-y')\right)^2 \left(\phi_{\rm MF}(x-y)\right)^2 
 \left(\int_0^L dz \,\phi^*_{\rm MF}(z-y')\phi_{\rm MF}(z-y)\right)^{N-2},\nonumber
\end{eqnarray}
where the last term under integral is
overlap between  product states of  orbitals $\phi_{\rm MF}$ occupied by ($N-2$) atoms, shifted by $(y-y')$. 
In the limit $N\to\infty$ the term $\mathcal{N}^2\left(\int_0^L dz \,\phi^*_{\rm MF}(z-y')\phi_{\rm MF}(z-y)\right)^{N-2}$ quickly decay at points $y \neq y'$. If this term was approximated by a delta function, precisely $L \delta(y-y')$, then the interaction energy \eqref{eq:appendix:ansatz-eint} would coincide with the interaction energy of a product of MF solitons. Yet, we keep this term and see that for our finite size system it makes a difference.

Similarly we evaluate the kinetic energy:
\begin{eqnarray}
E_{\rm kin} &=& \int_0^{L}\,{\rm d}x_1\ldots \int_0^{L}\,{\rm d}x_N\, \psi^*_{\rm Ansatz} (x_1,\,x_2,\,\ldots,x_N)
\bb{\sum_{j=1}^N \frac{-\hbar^2}{2m}\partial^2_{x_j} }
\psi_{\rm Ansatz} (x_1,\,x_2,\,\ldots,x_N)\\
 &=&\frac{-\hbar^2N}{2m}
 \int_0^{L}\,{\rm d}y
 \int_0^{L}\,{\rm d}y^{\prime}
  \,e^{i \frac{2\pi}{L}K (y-y')} 
  \,\bb{\int_0^{L}\,{\rm d}x_1
  \phi^*_{\rm MF}(x_1-y')\partial^2_{x_1}\phi_{\rm MF}(x_1-y)}
  \left(\int_0^L dz \,\phi^*_{\rm MF}(z-y')\phi_{\rm MF}(z-y)\right)^{N-1}
 ,\nonumber
\end{eqnarray}
where we used indistinguishability of bosons.

\subsection{Overlap between the state after $m$ measurements and yrast state\label{appendix:overlap-after-measurmenet}}

In Fig. \ref{fig:weights} we present the projection of the state $\ket{\tilde{\psi}^m_{\rm Ansatz}}$ on the subspace of yrast states. This projection is evaluated as $\sum_{K'=0}^{N-m} |a_{K'}|^2$, where 
\begin{equation}
    a_{K'} := \langle {K'}|\tilde{\psi}^m_{\rm Ansatz}\rangle
\end{equation}
is the overlap between a state after $m$ measurement and an  yrast state $\ket{K'}$ with $N-m$ atoms. We compute the values of $a_{K'}$ under an assumption that the Ansatz~\eqref{eq:appendix:smeared-sol} is a fair approximation of the yrast state.
Then $a_{K'}$ reads:
\begin{eqnarray}
    a_{K'} = \langle K'  | \tilde{\psi}^m_{\rm Ansatz}\rangle &=& \mathcal{N}_m \mathcal{N}_{K'}  \int_0^L\,dy\int_0^L\,dy' \,e^{i \frac{2\pi}{L} ( K y-K'y')} \bb{\prod_{j=1}^m \phi_{\rm MF; K}(x_j-y)} \nonumber\\
 & \times & \left(\int_0^L dz \phi^*_{\rm MF;K'}(z-y')\phi_{\rm MF;K}(z-y)\right)^{(N-m)},
\end{eqnarray}
where we introduce another notation for solitonic solution of NLSE~\eqref{eq:NLSE}, $\phi_{\rm MF; K}$ to indicate the momentum $2\pi\hbar K /(NL)$. The parameters $x_j$ are the random positions, at which particle were measured,  drawn from probability density function~\eqref{appendix:rho_m}, and $\mathcal{N}_m$, $\mathcal{N}_{K'}$ are the normalization factors of $\ket{\tilde{\psi}^m_{\rm Ansatz}}$ and $\ket{K'}$, respectively.

\subsection{Numerical methods\label{appendix:numerical-methods}}
We discuss our numerical methods on an example 
of the interaction energy Eq. \eqref{eq:appendix:ansatz-eint}. 
The computation requires evaluation of overlap integrals 
$A[w] := \int dz\,\phi^*_{\rm MF}(z) \, \phi_{\rm MF}(z-w)$ and $B[w]:=\int dz\,\bb{\phi^*_{\rm MF}(z) \, \phi_{\rm MF}(z-w)}^2$
which we evaluate for discrete values of shift $w$ and store in a computer memory before the main computation. Then the interaction energy is approximated by:
\begin{equation}
E_{\rm int } \approx  \frac{g\,N (N-1)\,\mathcal{N}^2}{2}\sum_y\sum_{y'} B[y-y'] \, \bb{D[y-y']}^{N-2} \, e^{i \frac{2\pi}{L}K (y-y')}\,\bb{\Delta y}^2,
\end{equation}
where integrals $\int dy$ were discretized to $\sum_y$ with discrete values of $y$ separated by $\Delta y$. 
The function $D[y-y']$ is equal to $A[y-y']$ for $y>y'$ and it is equal to $\bb{A[y'-y]}^*$ otherwise.
The results presented in Figs. \ref{fig:energies} and \ref{fig:validity} were evaluated on a numerical grid $1000$  points, i.e. with $\Delta y = 0.001$.
Already for the simplest integration scheme possible, based on the rectangle rule, we achieved  converging results. 

The results presented in other figures were not so sensitive to the numerical grid (as we were not interested in small differences of energies evaluated in different approaches), therefore we used numerical grid with $80$ points only, i.e. with $\Delta y = 0.0125$.

\section{The Lieb-Liniger equations \label{appendix:LLeqs}}
Any eigenstate of LL Hamiltonian~\eqref{eq:LLham} is clearly defined by a set of real numbers $p_j$, so-called quasi-momenta, satisfying (see Eq. (2.15) in~\cite{LiebLiniger1963}):
\begin{equation}
\label{fromlieb}
(-)^{N-1}e^{-i\frac{L}{\hbar}p_j}=\exp\left(-2i\sum^N_{l=1}{\rm arctan}\left(\frac{\hbar(p_j-p_l)}{mg}\right)\right),\quad \forall_{j\neq l}\,p_j\neq p_l,
\end{equation}
where total momentum and total energy of such state may be expressed respectively as $\sum_{j=1}^Np_j$ and $\sum_{j=1}^N\frac{p_j^2}{2m}$. After taking the logarithm of both sides of \eqref{fromlieb} and multiplying by $i$ we get:
\begin{equation}
\frac{L}{\hbar} p_j=2\pi I_j-2\sum^N_{l=1}{\rm arctan}\left(\frac{\hbar(p_j-p_l)}{mg}\right);\quad \forall_{l\neq j}I_j\neq I_l,\quad  I_l+\frac{N+1}{2}\in \mathbf{Z},
\end{equation}
where $I_j$'s are called Bethe quantum numbers which uniquely characterize the state of a system. 
It is worth noting that the total momentum can be expressed as $\frac{2\pi \hbar}{L}\sum_{j=1}^N I_j$. The ground state corresponds to $\{I_j\}_{j=1}^N$ satisfying $I_{j+1}-I_j=1$, $I_1=-I_N$, i.e.
\begin{eqnarray}
    & &\{I_j\}_{j=1}^N=\{-\frac{N-1}{2},-\frac{N-3}{2},...,-1,0,1,...,+\frac{N-1}{2}\} \text{ for odd } N, \nonumber\\
    & &\{I_j\}_{j=1}^N=\{-\frac{N-1}{2},-\frac{N-3}{2},...,-\frac{1}{2},+\frac{1}{2},...,+\frac{N-1}{2}\} \text{ for even } N \nonumber.
\end{eqnarray}
Excitations of the type-I (Bogoliubov states) and the type-II (yrast states), as well as the first excited state with total momentum $\frac{2\pi\hbar}{L}K$ (for $K\geq 2$), may be generated by increasing the value of appropriate $I_j$ by $K$:
\begin{itemize}
\item Bogoliubov states: $I'_N=I_N+K$
\item yrast states: $I'_{N+1-K}=I_{N+1-K}+K$
\item the first excited state: $I'_{N+2-K}=I_{N+2-K}+K$
\end{itemize}
One may check that in the limit $g\rightarrow 0$ the formula for the first excited states correctly reproduces the Fock state with $N-K+1$ particles in orbital with momentum $0$, $K-2$ particles with momentum $2\pi\hbar/L$ and one with $2\cdot 2\pi\hbar/L$, i.e. $\ket{n_0: N-K+1,n_{2\pi\hbar/L}: K-2,n_{2\cdot 2\pi\hbar/L}: 1}$.

\subsection{Numerical evaluation}
Choosing $L$, $\hbar^2$/$mL^2$ and $\hbar/L$ as the units of length, energy and momentum respectively, we get:
\begin{equation}
p_j=2\pi I_j-2\sum^N_{l=1}{\rm arctan}\left(\frac{p_j-p_l}{g}\right).
\end{equation}
For numerical convenience, instead of solving the above system of $N$ equations, we translate it to the problem of minimizing the function of $N$ variables:
\begin{equation}
\min_{p_1,...,p_N}\sum_{j=1}^N\left(p_j-2\pi I_j+2\sum^N_{l=1}{\rm arctan}\left(\frac{p_j-p_l}{g}\right)\right)^2,
\end{equation}
which may be straightforwardly solved numerically even for the number of particles $N$ on the order of $500$. Total energy of such state is given as:
\begin{equation}
E=\frac{1}{2}\sum_{j=1}^Np_j^2.
\end{equation}
\end{widetext}


\begin{thebibliography}{47}%
	\makeatletter
	\providecommand \@ifxundefined [1]{%
		\@ifx{#1\undefined}
	}%
	\providecommand \@ifnum [1]{%
		\ifnum #1\expandafter \@firstoftwo
		\else \expandafter \@secondoftwo
		\fi
	}%
	\providecommand \@ifx [1]{%
		\ifx #1\expandafter \@firstoftwo
		\else \expandafter \@secondoftwo
		\fi
	}%
	\providecommand \natexlab [1]{#1}%
	\providecommand \enquote  [1]{``#1''}%
	\providecommand \bibnamefont  [1]{#1}%
	\providecommand \bibfnamefont [1]{#1}%
	\providecommand \citenamefont [1]{#1}%
	\providecommand \href@noop [0]{\@secondoftwo}%
	\providecommand \href [0]{\begingroup \@sanitize@url \@href}%
	\providecommand \@href[1]{\@@startlink{#1}\@@href}%
	\providecommand \@@href[1]{\endgroup#1\@@endlink}%
	\providecommand \@sanitize@url [0]{\catcode `\\12\catcode `\$12\catcode
		`\&12\catcode `\#12\catcode `\^12\catcode `\_12\catcode `\%12\relax}%
	\providecommand \@@startlink[1]{}%
	\providecommand \@@endlink[0]{}%
	\providecommand \url  [0]{\begingroup\@sanitize@url \@url }%
	\providecommand \@url [1]{\endgroup\@href {#1}{\urlprefix }}%
	\providecommand \urlprefix  [0]{URL }%
	\providecommand \Eprint [0]{\href }%
	\providecommand \doibase [0]{http://dx.doi.org/}%
	\providecommand \selectlanguage [0]{\@gobble}%
	\providecommand \bibinfo  [0]{\@secondoftwo}%
	\providecommand \bibfield  [0]{\@secondoftwo}%
	\providecommand \translation [1]{[#1]}%
	\providecommand \BibitemOpen [0]{}%
	\providecommand \bibitemStop [0]{}%
	\providecommand \bibitemNoStop [0]{.\EOS\space}%
	\providecommand \EOS [0]{\spacefactor3000\relax}%
	\providecommand \BibitemShut  [1]{\csname bibitem#1\endcsname}%
	\let\auto@bib@innerbib\@empty
	\bibitem [{\citenamefont {Lieb}(1963)}]{Lieb1963}%
	\BibitemOpen
	\bibfield  {author} {\bibinfo {author} {\bibfnamefont {E.~H.}\ \bibnamefont
			{Lieb}},\ }\href {\doibase 10.1103/PhysRev.130.1616} {\bibfield  {journal}
		{\bibinfo  {journal} {Phys. Rev.}\ }\textbf {\bibinfo {volume} {130}},\
		\bibinfo {pages} {1616} (\bibinfo {year} {1963})}\BibitemShut {NoStop}%
	\bibitem [{\citenamefont {Lieb}\ and\ \citenamefont
		{Liniger}(1963)}]{LiebLiniger1963}%
	\BibitemOpen
	\bibfield  {author} {\bibinfo {author} {\bibfnamefont {E.~H.}\ \bibnamefont
			{Lieb}}\ and\ \bibinfo {author} {\bibfnamefont {W.}~\bibnamefont {Liniger}},\
	}\href {\doibase 10.1103/PhysRev.130.1605} {\bibfield  {journal} {\bibinfo
			{journal} {Phys. Rev.}\ }\textbf {\bibinfo {volume} {130}},\ \bibinfo {pages}
		{1605} (\bibinfo {year} {1963})}\BibitemShut {NoStop}%
	\bibitem [{\citenamefont {Gogolin}\ \emph {et~al.}(2004)\citenamefont
		{Gogolin}, \citenamefont {Nersesyan},\ and\ \citenamefont
		{Tsvelik}}]{Gogolin2004Dec}%
	\BibitemOpen
	\bibfield  {author} {\bibinfo {author} {\bibfnamefont {A.~O.}\ \bibnamefont
			{Gogolin}}, \bibinfo {author} {\bibfnamefont {A.~A.}\ \bibnamefont
			{Nersesyan}}, \ and\ \bibinfo {author} {\bibfnamefont {A.~M.}\ \bibnamefont
			{Tsvelik}},\ }\href
	{https://www.cambridge.org/pl/academic/subjects/physics/condensed-matter-physics-nanoscience-and-mesoscopic-physics/bosonization-and-strongly-correlated-systems?format=PB&isbn=9780521617192}
	{\bibfield  {journal} {\bibinfo  {journal} {Cambridge University Press}\ }
		(\bibinfo {year} {2004})}\BibitemShut {NoStop}%
	\bibitem [{\citenamefont {Jiang}\ \emph {et~al.}(2015)\citenamefont {Jiang},
		\citenamefont {Chen},\ and\ \citenamefont {Guan}}]{Jiang_2015}%
	\BibitemOpen
	\bibfield  {author} {\bibinfo {author} {\bibfnamefont {Y.-Z.}\ \bibnamefont
			{Jiang}}, \bibinfo {author} {\bibfnamefont {Y.-Y.}\ \bibnamefont {Chen}}, \
		and\ \bibinfo {author} {\bibfnamefont {X.-W.}\ \bibnamefont {Guan}},\ }\href
	{\doibase 10.1088/1674-1056/24/5/050311} {\bibfield  {journal} {\bibinfo
			{journal} {Chinese Physics B}\ }\textbf {\bibinfo {volume} {24}},\ \bibinfo
		{pages} {050311} (\bibinfo {year} {2015})}\BibitemShut {NoStop}%
	\bibitem [{\citenamefont {Lang}\ \emph {et~al.}(2017)\citenamefont {Lang},
		\citenamefont {Hekking},\ and\ \citenamefont {Minguzzi}}]{Lang2017}%
	\BibitemOpen
	\bibfield  {author} {\bibinfo {author} {\bibfnamefont {G.}~\bibnamefont
			{Lang}}, \bibinfo {author} {\bibfnamefont {F.}~\bibnamefont {Hekking}}, \
		and\ \bibinfo {author} {\bibfnamefont {A.}~\bibnamefont {Minguzzi}},\ }\href
	{\doibase 10.21468/SciPostPhys.3.1.003} {\bibfield  {journal} {\bibinfo
			{journal} {SciPost Phys.}\ }\textbf {\bibinfo {volume} {3}},\ \bibinfo
		{pages} {003} (\bibinfo {year} {2017})}\BibitemShut {NoStop}%
	\bibitem [{\citenamefont {Cazalilla}\ \emph {et~al.}(2011)\citenamefont
		{Cazalilla}, \citenamefont {Citro}, \citenamefont {Giamarchi}, \citenamefont
		{Orignac},\ and\ \citenamefont {Rigol}}]{Cazalilla2011}%
	\BibitemOpen
	\bibfield  {author} {\bibinfo {author} {\bibfnamefont {M.~A.}\ \bibnamefont
			{Cazalilla}}, \bibinfo {author} {\bibfnamefont {R.}~\bibnamefont {Citro}},
		\bibinfo {author} {\bibfnamefont {T.}~\bibnamefont {Giamarchi}}, \bibinfo
		{author} {\bibfnamefont {E.}~\bibnamefont {Orignac}}, \ and\ \bibinfo
		{author} {\bibfnamefont {M.}~\bibnamefont {Rigol}},\ }\href {\doibase
		10.1103/RevModPhys.83.1405} {\bibfield  {journal} {\bibinfo  {journal} {Rev.
				Mod. Phys.}\ }\textbf {\bibinfo {volume} {83}},\ \bibinfo {pages} {1405}
		(\bibinfo {year} {2011})}\BibitemShut {NoStop}%
	\bibitem [{\citenamefont {Sowi{\ifmmode\acute{n}\else\'{n}\fi}ski}\ and\
		\citenamefont
		{Garc{\ifmmode\acute{\imath}\else\'{\i}\fi}a-March}(2019)}]{Sowinski2019Sep}%
	\BibitemOpen
	\bibfield  {author} {\bibinfo {author} {\bibfnamefont {T.}~\bibnamefont
			{Sowi{\ifmmode\acute{n}\else\'{n}\fi}ski}}\ and\ \bibinfo {author}
		{\bibfnamefont {M.~{\ifmmode\acute{A}\else\'{A}\fi}.}\ \bibnamefont
			{Garc{\ifmmode\acute{\imath}\else\'{\i}\fi}a-March}},\ }\href {\doibase
		10.1088/1361-6633/ab3a80} {\bibfield  {journal} {\bibinfo  {journal} {Rep.
				Prog. Phys.}\ }\textbf {\bibinfo {volume} {82}},\ \bibinfo {pages} {104401}
		(\bibinfo {year} {2019})}\BibitemShut {NoStop}%
	\bibitem [{\citenamefont {Kivshar}\ and\ \citenamefont
		{Luther-Davies}(1998)}]{Kivshar1998May}%
	\BibitemOpen
	\bibfield  {author} {\bibinfo {author} {\bibfnamefont {Y.~S.}\ \bibnamefont
			{Kivshar}}\ and\ \bibinfo {author} {\bibfnamefont {B.}~\bibnamefont
			{Luther-Davies}},\ }\href {\doibase 10.1016/S0370-1573(97)00073-2} {\bibfield
		{journal} {\bibinfo  {journal} {Phys. Rep.}\ }\textbf {\bibinfo {volume}
			{298}},\ \bibinfo {pages} {81} (\bibinfo {year} {1998})}\BibitemShut
	{NoStop}%
	\bibitem [{\citenamefont {Gross}(1961)}]{Gross1961}%
	\BibitemOpen
	\bibfield  {author} {\bibinfo {author} {\bibfnamefont {E.~P.}\ \bibnamefont
			{Gross}},\ }\href {\doibase 10.1007/BF02731494} {\bibfield  {journal}
		{\bibinfo  {journal} {Il Nuovo Cimento (1955-1965)}\ }\textbf {\bibinfo
			{volume} {20}},\ \bibinfo {pages} {454} (\bibinfo {year} {1961})}\BibitemShut
	{NoStop}%
	\bibitem [{\citenamefont {Johnson}(1976)}]{Johnson1976}%
	\BibitemOpen
	\bibfield  {author} {\bibinfo {author} {\bibfnamefont {R.~S.}\ \bibnamefont
			{Johnson}},\ }\href {\doibase 10.1098/rspa.1976.0015} {\bibfield  {journal}
		{\bibinfo  {journal} {Proc. R. Soc. Lond. A.}\ }\textbf {\bibinfo {volume}
			{347}},\ \bibinfo {pages} {537} (\bibinfo {year} {1976})}\BibitemShut
	{NoStop}%
	\bibitem [{\citenamefont {Zakharov}\ and\ \citenamefont
		{Shabat}(1973)}]{zakharov73}%
	\BibitemOpen
	\bibfield  {author} {\bibinfo {author} {\bibfnamefont {V.~E.}\ \bibnamefont
			{Zakharov}}\ and\ \bibinfo {author} {\bibfnamefont {A.}~\bibnamefont
			{Shabat}},\ }\href@noop {} {\bibfield  {journal} {\bibinfo  {journal} {Zh.
				Eksp. Teor. Fiz.}\ }\textbf {\bibinfo {volume} {64}},\ \bibinfo {pages}
		{1627} (\bibinfo {year} {1973})}\BibitemShut {NoStop}%
	\bibitem [{\citenamefont {Burger}\ \emph {et~al.}(1999)\citenamefont {Burger},
		\citenamefont {Bongs}, \citenamefont {Dettmer}, \citenamefont {Ertmer},
		\citenamefont {Sengstock}, \citenamefont {Sanpera}, \citenamefont
		{Shlyapnikov},\ and\ \citenamefont {Lewenstein}}]{Burger1999}%
	\BibitemOpen
	\bibfield  {author} {\bibinfo {author} {\bibfnamefont {S.}~\bibnamefont
			{Burger}}, \bibinfo {author} {\bibfnamefont {K.}~\bibnamefont {Bongs}},
		\bibinfo {author} {\bibfnamefont {S.}~\bibnamefont {Dettmer}}, \bibinfo
		{author} {\bibfnamefont {W.}~\bibnamefont {Ertmer}}, \bibinfo {author}
		{\bibfnamefont {K.}~\bibnamefont {Sengstock}}, \bibinfo {author}
		{\bibfnamefont {A.}~\bibnamefont {Sanpera}}, \bibinfo {author} {\bibfnamefont
			{G.~V.}\ \bibnamefont {Shlyapnikov}}, \ and\ \bibinfo {author} {\bibfnamefont
			{M.}~\bibnamefont {Lewenstein}},\ }\href {\doibase
		10.1103/PhysRevLett.83.5198} {\bibfield  {journal} {\bibinfo  {journal}
			{Phys. Rev. Lett.}\ }\textbf {\bibinfo {volume} {83}},\ \bibinfo {pages}
		{5198} (\bibinfo {year} {1999})}\BibitemShut {NoStop}%
	\bibitem [{\citenamefont {Heidemann}\ \emph {et~al.}(2009)\citenamefont
		{Heidemann}, \citenamefont {Zhdanov}, \citenamefont {S\"utterlin},
		\citenamefont {Thomas},\ and\ \citenamefont {Morfill}}]{heidman2009}%
	\BibitemOpen
	\bibfield  {author} {\bibinfo {author} {\bibfnamefont {R.}~\bibnamefont
			{Heidemann}}, \bibinfo {author} {\bibfnamefont {S.}~\bibnamefont {Zhdanov}},
		\bibinfo {author} {\bibfnamefont {R.}~\bibnamefont {S\"utterlin}}, \bibinfo
		{author} {\bibfnamefont {H.~M.}\ \bibnamefont {Thomas}}, \ and\ \bibinfo
		{author} {\bibfnamefont {G.~E.}\ \bibnamefont {Morfill}},\ }\href {\doibase
		10.1103/PhysRevLett.102.135002} {\bibfield  {journal} {\bibinfo  {journal}
			{Phys. Rev. Lett.}\ }\textbf {\bibinfo {volume} {102}},\ \bibinfo {pages}
		{135002} (\bibinfo {year} {2009})}\BibitemShut {NoStop}%
	\bibitem [{\citenamefont {Chabchoub}\ \emph {et~al.}(2013)\citenamefont
		{Chabchoub}, \citenamefont {Kimmoun}, \citenamefont {Branger}, \citenamefont
		{Hoffmann}, \citenamefont {Proment}, \citenamefont {Onorato},\ and\
		\citenamefont {Akhmediev}}]{Chabchoub2013}%
	\BibitemOpen
	\bibfield  {author} {\bibinfo {author} {\bibfnamefont {A.}~\bibnamefont
			{Chabchoub}}, \bibinfo {author} {\bibfnamefont {O.}~\bibnamefont {Kimmoun}},
		\bibinfo {author} {\bibfnamefont {H.}~\bibnamefont {Branger}}, \bibinfo
		{author} {\bibfnamefont {N.}~\bibnamefont {Hoffmann}}, \bibinfo {author}
		{\bibfnamefont {D.}~\bibnamefont {Proment}}, \bibinfo {author} {\bibfnamefont
			{M.}~\bibnamefont {Onorato}}, \ and\ \bibinfo {author} {\bibfnamefont
			{N.}~\bibnamefont {Akhmediev}},\ }\href {\doibase
		10.1103/PhysRevLett.110.124101} {\bibfield  {journal} {\bibinfo  {journal}
			{Phys. Rev. Lett.}\ }\textbf {\bibinfo {volume} {110}},\ \bibinfo {pages}
		{124101} (\bibinfo {year} {2013})}\BibitemShut {NoStop}%
	\bibitem [{\citenamefont {Tong}\ \emph {et~al.}(2010)\citenamefont {Tong},
		\citenamefont {Wu}, \citenamefont {Carr},\ and\ \citenamefont
		{Kalinikos}}]{Tong2010}%
	\BibitemOpen
	\bibfield  {author} {\bibinfo {author} {\bibfnamefont {W.}~\bibnamefont
			{Tong}}, \bibinfo {author} {\bibfnamefont {M.}~\bibnamefont {Wu}}, \bibinfo
		{author} {\bibfnamefont {L.~D.}\ \bibnamefont {Carr}}, \ and\ \bibinfo
		{author} {\bibfnamefont {B.~A.}\ \bibnamefont {Kalinikos}},\ }\href {\doibase
		10.1103/PhysRevLett.104.037207} {\bibfield  {journal} {\bibinfo  {journal}
			{Phys. Rev. Lett.}\ }\textbf {\bibinfo {volume} {104}},\ \bibinfo {pages}
		{037207} (\bibinfo {year} {2010})}\BibitemShut {NoStop}%
	\bibitem [{\citenamefont {Frantzeskakis}(2010)}]{Frantzeskakis2010}%
	\BibitemOpen
	\bibfield  {author} {\bibinfo {author} {\bibfnamefont {D.~J.}\ \bibnamefont
			{Frantzeskakis}},\ }\href {\doibase 10.1088/1751-8113/43/21/213001}
	{\bibfield  {journal} {\bibinfo  {journal} {Journal of Physics A:
				Mathematical and Theoretical}\ }\textbf {\bibinfo {volume} {43}},\ \bibinfo
		{pages} {213001} (\bibinfo {year} {2010})}\BibitemShut {NoStop}%
	\bibitem [{\citenamefont {Kulish}\ \emph {et~al.}(1976)\citenamefont {Kulish},
		\citenamefont {Manakov},\ and\ \citenamefont {Faddeev}}]{Kulish1976}%
	\BibitemOpen
	\bibfield  {author} {\bibinfo {author} {\bibfnamefont {P.~P.}\ \bibnamefont
			{Kulish}}, \bibinfo {author} {\bibfnamefont {S.~V.}\ \bibnamefont {Manakov}},
		\ and\ \bibinfo {author} {\bibfnamefont {L.~D.}\ \bibnamefont {Faddeev}},\
	}\href {\doibase 10.1007/BF01028912} {\bibfield  {journal} {\bibinfo
			{journal} {Theoretical and Mathematical Physics}\ }\textbf {\bibinfo {volume}
			{28}},\ \bibinfo {pages} {615} (\bibinfo {year} {1976})}\BibitemShut
	{NoStop}%
	\bibitem [{\citenamefont {Ishikawa}\ and\ \citenamefont
		{Takayama}(1980)}]{ishikawa1980}%
	\BibitemOpen
	\bibfield  {author} {\bibinfo {author} {\bibfnamefont {M.}~\bibnamefont
			{Ishikawa}}\ and\ \bibinfo {author} {\bibfnamefont {H.}~\bibnamefont
			{Takayama}},\ }\href {\doibase 10.1143/JPSJ.49.1242} {\bibfield  {journal}
		{\bibinfo  {journal} {Journal of the Physical Society of Japan}\ }\textbf
		{\bibinfo {volume} {49}},\ \bibinfo {pages} {1242} (\bibinfo {year}
		{1980})},\ \Eprint
	{http://arxiv.org/abs/https://doi.org/10.1143/JPSJ.49.1242}
	{https://doi.org/10.1143/JPSJ.49.1242} \BibitemShut {NoStop}%
	\bibitem [{\citenamefont {Mottelson}(1999)}]{Mottelson1999}%
	\BibitemOpen
	\bibfield  {author} {\bibinfo {author} {\bibfnamefont {B.}~\bibnamefont
			{Mottelson}},\ }\href {\doibase 10.1103/PhysRevLett.83.2695} {\bibfield
		{journal} {\bibinfo  {journal} {Phys. Rev. Lett.}\ }\textbf {\bibinfo
			{volume} {83}},\ \bibinfo {pages} {2695} (\bibinfo {year}
		{1999})}\BibitemShut {NoStop}%
	\bibitem [{\citenamefont {Syrwid}\ and\ \citenamefont
		{Sacha}(2015)}]{syrwid2015}%
	\BibitemOpen
	\bibfield  {author} {\bibinfo {author} {\bibfnamefont {A.}~\bibnamefont
			{Syrwid}}\ and\ \bibinfo {author} {\bibfnamefont {K.}~\bibnamefont {Sacha}},\
	}\href@noop {} {\bibfield  {journal} {\bibinfo  {journal} {Physical Review
				A}\ }\textbf {\bibinfo {volume} {92}},\ \bibinfo {pages} {032110} (\bibinfo
		{year} {2015})}\BibitemShut {NoStop}%
	\bibitem [{\citenamefont {Syrwid}\ \emph {et~al.}(2016)\citenamefont {Syrwid},
		\citenamefont {Brewczyk}, \citenamefont {Gajda},\ and\ \citenamefont
		{Sacha}}]{syrwid2016}%
	\BibitemOpen
	\bibfield  {author} {\bibinfo {author} {\bibfnamefont {A.}~\bibnamefont
			{Syrwid}}, \bibinfo {author} {\bibfnamefont {M.}~\bibnamefont {Brewczyk}},
		\bibinfo {author} {\bibfnamefont {M.}~\bibnamefont {Gajda}}, \ and\ \bibinfo
		{author} {\bibfnamefont {K.}~\bibnamefont {Sacha}},\ }\href@noop {}
	{\bibfield  {journal} {\bibinfo  {journal} {Physical Review A}\ }\textbf
		{\bibinfo {volume} {94}},\ \bibinfo {pages} {023623} (\bibinfo {year}
		{2016})}\BibitemShut {NoStop}%
	\bibitem [{\citenamefont {Sato}\ \emph {et~al.}(2012)\citenamefont {Sato},
		\citenamefont {Kanamoto}, \citenamefont {Kaminishi},\ and\ \citenamefont
		{Deguchi}}]{Sato2012}%
	\BibitemOpen
	\bibfield  {author} {\bibinfo {author} {\bibfnamefont {J.}~\bibnamefont
			{Sato}}, \bibinfo {author} {\bibfnamefont {R.}~\bibnamefont {Kanamoto}},
		\bibinfo {author} {\bibfnamefont {E.}~\bibnamefont {Kaminishi}}, \ and\
		\bibinfo {author} {\bibfnamefont {T.}~\bibnamefont {Deguchi}},\ }\href
	{\doibase 10.1103/PhysRevLett.108.110401} {\bibfield  {journal} {\bibinfo
			{journal} {Phys. Rev. Lett.}\ }\textbf {\bibinfo {volume} {108}},\ \bibinfo
		{pages} {110401} (\bibinfo {year} {2012})}\BibitemShut {NoStop}%
	\bibitem [{\citenamefont {Sato}\ \emph {et~al.}(2016)\citenamefont {Sato},
		\citenamefont {Kanamoto}, \citenamefont {Kaminishi},\ and\ \citenamefont
		{Deguchi}}]{Sato2016}%
	\BibitemOpen
	\bibfield  {author} {\bibinfo {author} {\bibfnamefont {J.}~\bibnamefont
			{Sato}}, \bibinfo {author} {\bibfnamefont {R.}~\bibnamefont {Kanamoto}},
		\bibinfo {author} {\bibfnamefont {E.}~\bibnamefont {Kaminishi}}, \ and\
		\bibinfo {author} {\bibfnamefont {T.}~\bibnamefont {Deguchi}},\ }\href
	{\doibase 10.1088/1367-2630/18/7/075008} {\bibfield  {journal} {\bibinfo
			{journal} {New Journal of Physics}\ }\textbf {\bibinfo {volume} {18}},\
		\bibinfo {pages} {075008} (\bibinfo {year} {2016})}\BibitemShut {NoStop}%
	\bibitem [{\citenamefont {Kaminishi}\ \emph {et~al.}(2018)\citenamefont
		{Kaminishi}, \citenamefont {Mori},\ and\ \citenamefont
		{Miyashita}}]{Kaminishi2019}%
	\BibitemOpen
	\bibfield  {author} {\bibinfo {author} {\bibfnamefont {E.}~\bibnamefont
			{Kaminishi}}, \bibinfo {author} {\bibfnamefont {T.}~\bibnamefont {Mori}}, \
		and\ \bibinfo {author} {\bibfnamefont {S.}~\bibnamefont {Miyashita}},\
	}\href@noop {} {\enquote {\bibinfo {title} {Construction of quantum dark
				soliton in one-dimensional bose gas},}\ } (\bibinfo {year} {2018}),\ \Eprint
	{http://arxiv.org/abs/arXiv:1811.00211} {arXiv:1811.00211} \BibitemShut
	{NoStop}%
	\bibitem [{\citenamefont {Shamailov}\ and\ \citenamefont
		{Brand}(2019)}]{Brand2019}%
	\BibitemOpen
	\bibfield  {author} {\bibinfo {author} {\bibfnamefont {S.~S.}\ \bibnamefont
			{Shamailov}}\ and\ \bibinfo {author} {\bibfnamefont {J.}~\bibnamefont
			{Brand}},\ }\href {\doibase 10.1103/PhysRevA.99.043632} {\bibfield  {journal}
		{\bibinfo  {journal} {Phys. Rev. A}\ }\textbf {\bibinfo {volume} {99}},\
		\bibinfo {pages} {043632} (\bibinfo {year} {2019})}\BibitemShut {NoStop}%
	\bibitem [{\citenamefont {Martin}\ and\ \citenamefont
		{Ruostekoski}(2010)}]{martin2010prl}%
	\BibitemOpen
	\bibfield  {author} {\bibinfo {author} {\bibfnamefont {A.~D.}\ \bibnamefont
			{Martin}}\ and\ \bibinfo {author} {\bibfnamefont {J.}~\bibnamefont
			{Ruostekoski}},\ }\href {\doibase 10.1103/PhysRevLett.104.194102} {\bibfield
		{journal} {\bibinfo  {journal} {Phys. Rev. Lett.}\ }\textbf {\bibinfo
			{volume} {104}},\ \bibinfo {pages} {194102} (\bibinfo {year}
		{2010})}\BibitemShut {NoStop}%
	\bibitem [{\citenamefont {Katsimiga}\ \emph
		{et~al.}(2017{\natexlab{a}})\citenamefont {Katsimiga}, \citenamefont
		{Mistakidis}, \citenamefont {Koutentakis}, \citenamefont {Kevrekidis},\ and\
		\citenamefont {Schmelcher}}]{Katsimiga2017bent}%
	\BibitemOpen
	\bibfield  {author} {\bibinfo {author} {\bibfnamefont {G.~C.}\ \bibnamefont
			{Katsimiga}}, \bibinfo {author} {\bibfnamefont {S.~I.}\ \bibnamefont
			{Mistakidis}}, \bibinfo {author} {\bibfnamefont {G.~M.}\ \bibnamefont
			{Koutentakis}}, \bibinfo {author} {\bibfnamefont {P.~G.}\ \bibnamefont
			{Kevrekidis}}, \ and\ \bibinfo {author} {\bibfnamefont {P.}~\bibnamefont
			{Schmelcher}},\ }\href {\doibase 10.1088/1367-2630/aa96f6} {\bibfield
		{journal} {\bibinfo  {journal} {New Journal of Physics}\ }\textbf {\bibinfo
			{volume} {19}},\ \bibinfo {pages} {123012} (\bibinfo {year}
		{2017}{\natexlab{a}})}\BibitemShut {NoStop}%
	\bibitem [{\citenamefont {Katsimiga}\ \emph {et~al.}(2018)\citenamefont
		{Katsimiga}, \citenamefont {Mistakidis}, \citenamefont {Koutentakis},
		\citenamefont {Kevrekidis},\ and\ \citenamefont
		{Schmelcher}}]{Katsimiga2018}%
	\BibitemOpen
	\bibfield  {author} {\bibinfo {author} {\bibfnamefont {G.~C.}\ \bibnamefont
			{Katsimiga}}, \bibinfo {author} {\bibfnamefont {S.~I.}\ \bibnamefont
			{Mistakidis}}, \bibinfo {author} {\bibfnamefont {G.~M.}\ \bibnamefont
			{Koutentakis}}, \bibinfo {author} {\bibfnamefont {P.~G.}\ \bibnamefont
			{Kevrekidis}}, \ and\ \bibinfo {author} {\bibfnamefont {P.}~\bibnamefont
			{Schmelcher}},\ }\href {\doibase 10.1103/PhysRevA.98.013632} {\bibfield
		{journal} {\bibinfo  {journal} {Phys. Rev. A}\ }\textbf {\bibinfo {volume}
			{98}},\ \bibinfo {pages} {013632} (\bibinfo {year} {2018})}\BibitemShut
	{NoStop}%
	\bibitem [{\citenamefont {Delande}\ and\ \citenamefont
		{Sacha}(2014)}]{Delande2014}%
	\BibitemOpen
	\bibfield  {author} {\bibinfo {author} {\bibfnamefont {D.}~\bibnamefont
			{Delande}}\ and\ \bibinfo {author} {\bibfnamefont {K.}~\bibnamefont
			{Sacha}},\ }\href {\doibase 10.1103/PhysRevLett.112.040402} {\bibfield
		{journal} {\bibinfo  {journal} {Phys. Rev. Lett.}\ }\textbf {\bibinfo
			{volume} {112}},\ \bibinfo {pages} {040402} (\bibinfo {year}
		{2014})}\BibitemShut {NoStop}%
	\bibitem [{Note1()}]{Note1}%
	\BibitemOpen
	\bibinfo {note} {We rescaled the original Lieb-Liniger model by a factor $2$,
		to have the form which is used more frequently now.}\BibitemShut {Stop}%
	\bibitem [{\citenamefont {DeWitt}(1958)}]{deWitt1958}%
	\BibitemOpen
	\bibfield  {author} {\bibinfo {author} {\bibfnamefont {C.}~\bibnamefont
			{DeWitt}},\ }\href@noop {} {\emph {\bibinfo {title} {The Many Body
				Problem}}}\ (\bibinfo  {publisher} {John Wiley R Sons, Inc.},\ \bibinfo
	{year} {1958})\BibitemShut {NoStop}%
	\bibitem [{\citenamefont {O\l{}dziejewski}\ \emph {et~al.}(2018)\citenamefont
		{O\l{}dziejewski}, \citenamefont {G\'orecki}, \citenamefont {Paw\l{}owski},\
		and\ \citenamefont {Rz\k{a}\ifmmode~\dot{z}\else
			\.{z}\fi{}ewski}}]{rafal2018idealSol}%
	\BibitemOpen
	\bibfield  {author} {\bibinfo {author} {\bibfnamefont {R.}~\bibnamefont
			{O\l{}dziejewski}}, \bibinfo {author} {\bibfnamefont {W.}~\bibnamefont
			{G\'orecki}}, \bibinfo {author} {\bibfnamefont {K.}~\bibnamefont
			{Paw\l{}owski}}, \ and\ \bibinfo {author} {\bibfnamefont {K.}~\bibnamefont
			{Rz\k{a}\ifmmode~\dot{z}\else \.{z}\fi{}ewski}},\ }\href {\doibase
		10.1103/PhysRevA.97.063617} {\bibfield  {journal} {\bibinfo  {journal} {Phys.
				Rev. A}\ }\textbf {\bibinfo {volume} {97}},\ \bibinfo {pages} {063617}
		(\bibinfo {year} {2018})}\BibitemShut {NoStop}%
	\bibitem [{\citenamefont {Kanamoto}\ \emph {et~al.}(2008)\citenamefont
		{Kanamoto}, \citenamefont {Carr},\ and\ \citenamefont
		{Ueda}}]{KanamotoCarr2008}%
	\BibitemOpen
	\bibfield  {author} {\bibinfo {author} {\bibfnamefont {R.}~\bibnamefont
			{Kanamoto}}, \bibinfo {author} {\bibfnamefont {L.~D.}\ \bibnamefont {Carr}},
		\ and\ \bibinfo {author} {\bibfnamefont {M.}~\bibnamefont {Ueda}},\ }\href
	{\doibase 10.1103/PhysRevLett.100.060401} {\bibfield  {journal} {\bibinfo
			{journal} {Phys. Rev. Lett.}\ }\textbf {\bibinfo {volume} {100}},\ \bibinfo
		{pages} {060401} (\bibinfo {year} {2008})}\BibitemShut {NoStop}%
	\bibitem [{\citenamefont {Kaminishi}\ \emph {et~al.}(2011)\citenamefont
		{Kaminishi}, \citenamefont {Kanamoto}, \citenamefont {Sato},\ and\
		\citenamefont {Deguchi}}]{Kaminishi2011}%
	\BibitemOpen
	\bibfield  {author} {\bibinfo {author} {\bibfnamefont {E.}~\bibnamefont
			{Kaminishi}}, \bibinfo {author} {\bibfnamefont {R.}~\bibnamefont {Kanamoto}},
		\bibinfo {author} {\bibfnamefont {J.}~\bibnamefont {Sato}}, \ and\ \bibinfo
		{author} {\bibfnamefont {T.}~\bibnamefont {Deguchi}},\ }\href {\doibase
		10.1103/PhysRevA.83.031601} {\bibfield  {journal} {\bibinfo  {journal} {Phys.
				Rev. A}\ }\textbf {\bibinfo {volume} {83}},\ \bibinfo {pages} {031601}
		(\bibinfo {year} {2011})}\BibitemShut {NoStop}%
	\bibitem [{\citenamefont {Fialko}\ \emph {et~al.}(2012)\citenamefont {Fialko},
		\citenamefont {Delattre}, \citenamefont {Brand},\ and\ \citenamefont
		{Kolovsky}}]{Fialko2012}%
	\BibitemOpen
	\bibfield  {author} {\bibinfo {author} {\bibfnamefont {O.}~\bibnamefont
			{Fialko}}, \bibinfo {author} {\bibfnamefont {M.-C.}\ \bibnamefont
			{Delattre}}, \bibinfo {author} {\bibfnamefont {J.}~\bibnamefont {Brand}}, \
		and\ \bibinfo {author} {\bibfnamefont {A.~R.}\ \bibnamefont {Kolovsky}},\
	}\href {\doibase 10.1103/PhysRevLett.108.250402} {\bibfield  {journal}
		{\bibinfo  {journal} {Phys. Rev. Lett.}\ }\textbf {\bibinfo {volume} {108}},\
		\bibinfo {pages} {250402} (\bibinfo {year} {2012})}\BibitemShut {NoStop}%
	\bibitem [{\citenamefont {O{\l}dziejewski}\ \emph {et~al.}(2018)\citenamefont
		{O{\l}dziejewski}, \citenamefont {G{\'{o}}recki}, \citenamefont
		{Paw{\l}owski},\ and\ \citenamefont
		{Rz{\k{a}}{\.{z}}ewski}}]{rafal2018roton}%
	\BibitemOpen
	\bibfield  {author} {\bibinfo {author} {\bibfnamefont {R.}~\bibnamefont
			{O{\l}dziejewski}}, \bibinfo {author} {\bibfnamefont {W.}~\bibnamefont
			{G{\'{o}}recki}}, \bibinfo {author} {\bibfnamefont {K.}~\bibnamefont
			{Paw{\l}owski}}, \ and\ \bibinfo {author} {\bibfnamefont {K.}~\bibnamefont
			{Rz{\k{a}}{\.{z}}ewski}},\ }\href {\doibase 10.1088/1367-2630/aaf295}
	{\bibfield  {journal} {\bibinfo  {journal} {New Journal of Physics}\ }\textbf
		{\bibinfo {volume} {20}},\ \bibinfo {pages} {123006} (\bibinfo {year}
		{2018})}\BibitemShut {NoStop}%
	\bibitem [{\citenamefont {Katsimiga}\ \emph
		{et~al.}(2017{\natexlab{b}})\citenamefont {Katsimiga}, \citenamefont
		{Koutentakis}, \citenamefont {Mistakidis}, \citenamefont {Kevrekidis},\ and\
		\citenamefont {Schmelcher}}]{Katsimiga2017db}%
	\BibitemOpen
	\bibfield  {author} {\bibinfo {author} {\bibfnamefont {G.~C.}\ \bibnamefont
			{Katsimiga}}, \bibinfo {author} {\bibfnamefont {G.~M.}\ \bibnamefont
			{Koutentakis}}, \bibinfo {author} {\bibfnamefont {S.~I.}\ \bibnamefont
			{Mistakidis}}, \bibinfo {author} {\bibfnamefont {P.~G.}\ \bibnamefont
			{Kevrekidis}}, \ and\ \bibinfo {author} {\bibfnamefont {P.}~\bibnamefont
			{Schmelcher}},\ }\href {\doibase 10.1088/1367-2630/aa766b} {\bibfield
		{journal} {\bibinfo  {journal} {New Journal of Physics}\ }\textbf {\bibinfo
			{volume} {19}},\ \bibinfo {pages} {073004} (\bibinfo {year}
		{2017}{\natexlab{b}})}\BibitemShut {NoStop}%
	\bibitem [{\citenamefont {Mistakidis}\ \emph {et~al.}(2018)\citenamefont
		{Mistakidis}, \citenamefont {Katsimiga}, \citenamefont {Kevrekidis},\ and\
		\citenamefont {Schmelcher}}]{Mistakidis2018}%
	\BibitemOpen
	\bibfield  {author} {\bibinfo {author} {\bibfnamefont {S.~I.}\ \bibnamefont
			{Mistakidis}}, \bibinfo {author} {\bibfnamefont {G.~C.}\ \bibnamefont
			{Katsimiga}}, \bibinfo {author} {\bibfnamefont {P.~G.}\ \bibnamefont
			{Kevrekidis}}, \ and\ \bibinfo {author} {\bibfnamefont {P.}~\bibnamefont
			{Schmelcher}},\ }\href {\doibase 10.1088/1367-2630/aabc6a} {\bibfield
		{journal} {\bibinfo  {journal} {New Journal of Physics}\ }\textbf {\bibinfo
			{volume} {20}},\ \bibinfo {pages} {043052} (\bibinfo {year}
		{2018})}\BibitemShut {NoStop}%
	\bibitem [{\citenamefont {Castin}\ and\ \citenamefont
		{Herzog}(2001)}]{Castin2001}%
	\BibitemOpen
	\bibfield  {author} {\bibinfo {author} {\bibfnamefont {Y.}~\bibnamefont
			{Castin}}\ and\ \bibinfo {author} {\bibfnamefont {C.}~\bibnamefont
			{Herzog}},\ }\href {\doibase 10.1016/S1296-2147(01)01183-0} {\bibfield
		{journal} {\bibinfo  {journal} {Comptes Rendus de l'Academie des Sciences de
				Paris}\ }\textbf {\bibinfo {volume} {2}},\ \bibinfo {pages} {419} (\bibinfo
		{year} {2001})}\BibitemShut {NoStop}%
	\bibitem [{\citenamefont {Penrose}\ and\ \citenamefont
		{Onsager}(1956)}]{Penrose1956}%
	\BibitemOpen
	\bibfield  {author} {\bibinfo {author} {\bibfnamefont {O.}~\bibnamefont
			{Penrose}}\ and\ \bibinfo {author} {\bibfnamefont {L.}~\bibnamefont
			{Onsager}},\ }\href {\doibase 10.1103/PhysRev.104.576} {\bibfield  {journal}
		{\bibinfo  {journal} {Phys. Rev.}\ }\textbf {\bibinfo {volume} {104}},\
		\bibinfo {pages} {576} (\bibinfo {year} {1956})}\BibitemShut {NoStop}%
	\bibitem [{Note2()}]{Note2}%
	\BibitemOpen
	\bibinfo {note} {For details of our calculations see Appendix \ref
		{appendix:computation-state}.}\BibitemShut {Stop}%
	\bibitem [{Note3()}]{Note3}%
	\BibitemOpen
	\bibinfo {note} {For details of our computations see Appendix \ref
		{appendix:overlap-after-measurmenet}}\BibitemShut {NoStop}%
	\bibitem [{\citenamefont {Carr}\ \emph {et~al.}(2000)\citenamefont {Carr},
		\citenamefont {Clark},\ and\ \citenamefont {Reinhardt}}]{Carr2000}%
	\BibitemOpen
	\bibfield  {author} {\bibinfo {author} {\bibfnamefont {L.~D.}\ \bibnamefont
			{Carr}}, \bibinfo {author} {\bibfnamefont {C.~W.}\ \bibnamefont {Clark}}, \
		and\ \bibinfo {author} {\bibfnamefont {W.~P.}\ \bibnamefont {Reinhardt}},\
	}\href {\doibase 10.1103/PhysRevA.62.063611} {\bibfield  {journal} {\bibinfo
			{journal} {Phys. Rev. A}\ }\textbf {\bibinfo {volume} {62}},\ \bibinfo
		{pages} {063611} (\bibinfo {year} {2000})}\BibitemShut {NoStop}%
	\bibitem [{Note4()}]{Note4}%
	\BibitemOpen
	\bibinfo {note} {Private communication.}\BibitemShut {Stop}%
	\bibitem [{\citenamefont {Paw\l{}owski}\ and\ \citenamefont
		{Rza\.zewski}(2015)}]{dipoloweSolitony2015}%
	\BibitemOpen
	\bibfield  {author} {\bibinfo {author} {\bibfnamefont {K.}~\bibnamefont
			{Paw\l{}owski}}\ and\ \bibinfo {author} {\bibfnamefont {K.}~\bibnamefont
			{Rza\.zewski}},\ }\href {http://stacks.iop.org/1367-2630/17/i=10/a=105006}
	{\bibfield  {journal} {\bibinfo  {journal} {New Journal of Physics}\ }\textbf
		{\bibinfo {volume} {17}},\ \bibinfo {pages} {105006} (\bibinfo {year}
		{2015})}\BibitemShut {NoStop}%
	\bibitem [{\citenamefont {Reichert}\ \emph {et~al.}(2019)\citenamefont
		{Reichert}, \citenamefont {Astrakharchik}, \citenamefont
		{Petkovi\ifmmode~\acute{c}\else \'{c}\fi{}},\ and\ \citenamefont
		{Ristivojevic}}]{Reichert2019}%
	\BibitemOpen
	\bibfield  {author} {\bibinfo {author} {\bibfnamefont {B.}~\bibnamefont
			{Reichert}}, \bibinfo {author} {\bibfnamefont {G.~E.}\ \bibnamefont
			{Astrakharchik}}, \bibinfo {author} {\bibfnamefont {A.}~\bibnamefont
			{Petkovi\ifmmode~\acute{c}\else \'{c}\fi{}}}, \ and\ \bibinfo {author}
		{\bibfnamefont {Z.}~\bibnamefont {Ristivojevic}},\ }\href {\doibase
		10.1103/PhysRevLett.123.250602} {\bibfield  {journal} {\bibinfo  {journal}
			{Phys. Rev. Lett.}\ }\textbf {\bibinfo {volume} {123}},\ \bibinfo {pages}
		{250602} (\bibinfo {year} {2019})}\BibitemShut {NoStop}%
	\bibitem [{\citenamefont {Gaudin}(1971)}]{Gaudin1971}%
	\BibitemOpen
	\bibfield  {author} {\bibinfo {author} {\bibfnamefont {M.}~\bibnamefont
			{Gaudin}},\ }\href {\doibase 10.1103/PhysRevA.4.386} {\bibfield  {journal}
		{\bibinfo  {journal} {Phys. Rev. A}\ }\textbf {\bibinfo {volume} {4}},\
		\bibinfo {pages} {386} (\bibinfo {year} {1971})}\BibitemShut {NoStop}%
\end{thebibliography}
\end{document}